\documentclass[a4paper,fleqn,usenatbib,useAMS]{mnras}

\usepackage{graphicx}	
\usepackage{amsmath}	
\usepackage{amssymb}	
\usepackage{multicol}        
\usepackage{bm}		
\usepackage{pdflscape}	

\usepackage[T1]{fontenc}
\usepackage{ae,aecompl}

\usepackage{newtxtext,newtxmath}

\title[Dangerous dust clouds above lunar surface]{\textit{Dangerous dust clouds above lunar surface}}

\author[O. V. Arkhypov et al.]{Oleksiy V. Arkhypov, Maxim L. Khodachenko 
\thanks{Contact e-mail: \href{mailto:maxim.khodachenko@oeaw.ac.at}{maxim.khodachenko@oeaw.ac.at}}
\\
Space Research Institute, Austrian Academy of Sciences, Schmiedlstrasse 6, 8042 Graz, Austria \\
}

\date{Last updated 2023; in original form 2023}

\pubyear{2023}

\begin{document}
\label{firstpage}
\pagerange{\pageref{firstpage}--\pageref{lastpage}}
\maketitle

\begin{abstract}
It is believed that the levitating lunar dust was discovered during the lunar missions of 1960s. Since that time the dust component of lunar exosphere is a subject of intensive investigations, based on the data from these, and especially, from the recent missions, which at the same time cover a rather limited time-frame of the mission ($\lesssim$ 1 yr). Moreover, the performed studies give highly controversial results regarding dust concentration.
At the same time, the documented information provided by the Earth-based monitoring of the Moon during at least the last three centuries still remains unused. In the present study, we fill this data analysis gap.
The survey of historical reports of the XVIII-XX centuries devoted to observational manifestations of the supposed lunar atmosphere and comparing them with the modern data from space missions, as well as with the available data on hundreds of anomalous (i.e. too long-lasting) stellar occultations by the lunar limb, enable us revealing numerous evidences of the lunar dust phenomena. By  modeling of the conditions of such observations, we determine the Sun-Moon elongation angles and calculate the altitudes as well as phase angles of the dust clouds, which scattered the sunlight during the particular events. Using this information, as well as the Mie scattering theory, we estimate the concentration of dust and its damaging effect at different orbits of a possible spacecraft.
It was found that the dust clouds might have different altitudes and space scales, ranging from several kilometers up to the whole-Moon global formations manifested as an "annular Moon". The latter demonstrate dust presence at the altitudes of up to hundreds kilometers, which is much higher than the typically known values. Apparently, the rarity of such phenomena is a reason for their in-situ non-detection by the space missions with a limited time-frame of operation. Our estimates show that these global dust clouds of sub-$\mu\textrm{m}$ grains could crash a space-vehicle at the low ($<10$) km orbits, similar to the incidents with landers \textit{Vikram}, \textit{Beresheet} and \textit{Hakuto-R M1}. The statistics of dust clouds' appearance enabled a reconstruction of a typical shape of a local dust cloud which resembles the shape of an impact plume. This, together with the revealed seasonal periodicity of observational manifestations of the dust phenomena, confirm a hypothesis on the meteoroid impact nature of the majority of the circumlunar dust clouds. At the same time, the discovered additional periodicity of the dust cloud appearance at half of synodic lunar month ($14.77$ d), as well as the specific dust activity regions identified on the lunar surface, argue for an additional non-impact source of the circumlunar dust, connected with the lunar outgassing events, controlled by the solar tidal effects, completely unstudied. Moreover, the tendency of dust clouds to be observed during the low-level solar activity raises a question on possible dust pick up by the solar wind flow.
The obtained results are important with respect to the growing technological activity around and on the Moon, including the plans of landers and manned missions. The "forgotten" old Earth-based observations combined with the available more recent records of other observational manifestations of the circumlunar dust shed the light on this still unstudied, but dangerous phenomenon on the Moon, which should be seriously considered in the context of modern space missions.
\end{abstract}





\section{Introduction}

Space probes of the Surveyor series (5th, 6th and 7th), which performed soft landings on the Moon, registered a line-of-light stretching along the lunar horizon after sunsets. It was found that this glow near the Moon horizon is a sunlight scattered by dust particles with a diameter of an order of 10 $\mu$m, which formed a temporary cloud that lasted up to 3 hours in the terminator region (Rennilson and Criswell 1974).

The phenomenon of lunar limb glow was confirmed by astronauts flying around the Moon (McCoy 1974). An analysis of their sketches and photographs showed that fine lunar dust (of an order of a $\mu$m fraction in diameter) reached heights of more than 100 km and had a mass of about 10 kg per square kilometer (McCoy 1976). The onboard photometry of the rover \textit{Lunokhod-2} also registered the phenomenon of lunar twilight caused by the scattering of sunlight by dust at an altitude of more than half a kilometer (McCoy 1976).
However, the later \textit{Clementine} and \textit{Lunar Reconnaissance Orbiter} missions found that the concentration of lunar dust above the lunar surface appears $10^4$ times lower than that followed from the Apollo estimates (Hor\'{a}nyi et al. 2015).

Finally, in 2013, a specialized mission, the \textit{Lunar Atmosphere and Dust Environment Explorer} (LADEE), was launched on a low lunar orbit. Its dust counter registered 5 dust clouds at the altitude of only 75 km, which were located at the terminator region and had a density comparable to the estimates based on Apollo data (Lianghai et al. 2020). It was argued that "\textit{the so-called dust exosphere is not a global phenomenon but just a local electrified dust fountain near twilight craters}". At the same time, there are valuable data of numerous Earth-based observations of the Moon accumulated during the previous 300 years, which attest the existence of dense and sometimes global lunar dust clouds. Nevertheless, there is a tradition to refer only to the data from space missions in all publications dedicated to the levitating lunar dust (e.g., Horay et al. 2015), whereas the old observations remain ignored. In the present paper, we fill this gap and include into the scope of analysis the observational material and historical reports from 18th-19th centuries up to recent times. Altogether, the information about dusty phenomena in the close vicinity of the Moon is important for assessing the risk of spacecraft and landers in the circumlunar environment. This is especially actual in connection with the accidents of many space probes (e.g., \textit{Surveyor 4}, \textit{Luna 11, 15, 18, 23}; \textit{Hagoromo}, \textit{Longjiang 1}, \textit{Beresheet}, \textit{Vikram} and Hakuto-R M1) as well as with the existing plans for the future manned flights to the Moon.

\section{Remote sensing of dust clouds}

Depending on the type of light signature, various manifestations of the levitating dust are possible in the circumlunar space.
We consider different theoretical scenarios for the observed optical phenomena, which can be used for detection and remote sensing of the dust formations near the Moon.

\subsection{Direct light extinction phenomena}

Predictably, the phenomenology of extinction phenomena depends on the angular size of the light source. That is why the description of such appearances must be differentiated according to the type of sources.

\subsubsection{Dark band of planetary occultations}

\begin{figure}
   \centering
   \includegraphics[width=0.4\textwidth,clip=]{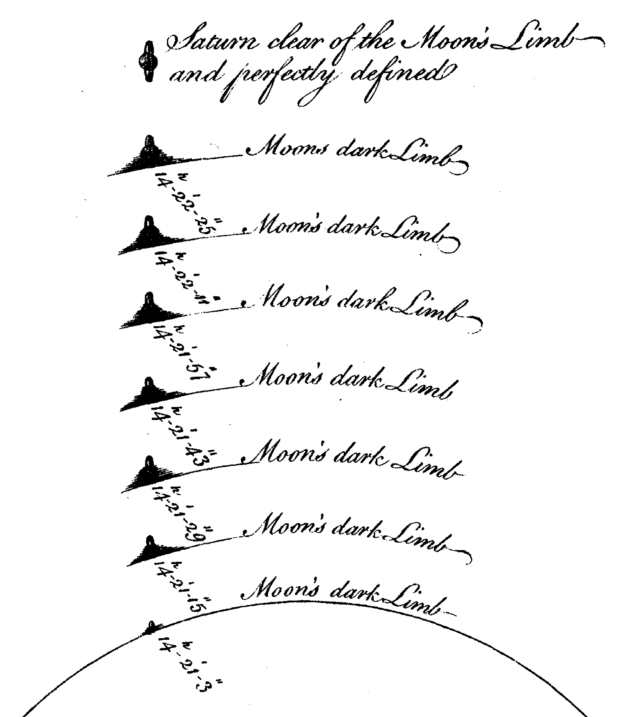}
   \caption{Sketch of a hazy view of the Saturn during its emergence from behind the dark side of the Moon on 17 June, 1762 (Dunn 1762).}
              \label{Fig01}
    \end{figure}

Apparently, the first documented manifestation of the absorption of light in a lunar dust cloud is the anomalous appearance of Saturn as it emerged from behind the Moon on June 17, 1762 (Dunn 1762). The observer, Samuel Dunn from Chelsea in London, described the phenomenon (Fig.~\ref{Fig01}) as follows:

\textit{"At 14h 21m 3s, I saw a faint point of light, where the emersion afterwards appeared ... I judged it was the tip of the ring just emerging. Yet it appeared so dull and hazy, that I had suspected my telescope, if I had not known it to have been rightly adjusted. At 14h 22m 4s, the preceding part of the ring was emerged, and it appeared more bright; and now the body seemed emerging or emerged, but so very hazy and ill defined, both the body and the ring confused together, closely on the Moon's dark limb, that I should not have taken it for Saturn, but for a comet emerging from behind the Moon, had I not known otherwise from the tables, or seen Saturn the preceding mornings. At 14h 22m 30s, the preceding end of the ring more plain and bright, the subsequent end of the ring more dull; and the body, at this time, appeared a little more distinct than before...
At 14h 22m 34s, the subsequent end of the ring appeared most dull, and the preceding end clear; after which, in some short space of time, the whole ring and body of Saturn appeared sharply and well defined. Wherefore, I conclude, that this diversity of appearance must have arisen from the effects of an atmosphere of the Moon."} (Dunn 1762).

Similar effects (phenomena) were seen repeatedly, during the occultations of Saturn and Jupiter by the Moon, as well as during the occultation of Mars (see in Table 1). In particular, different observers independently reported that "\textit{a darkish line}" (Goddard 1856) was seen, or "\textit{a dark line was plainly perceptible}" (Grove 1856) in parallel with the lunar limb on the Jupiter background during the grazing occultation on November 7, 1856. One of the observers explained: "\textit{The dark line I thought at first arose from the intervention of a fine rim of the dark body of the Moon, but this could hardly be the case towards the termination of the occultation}" (Grove 1856).

\begin{figure}
   \centering
   \includegraphics[width=0.375\textwidth,clip=]{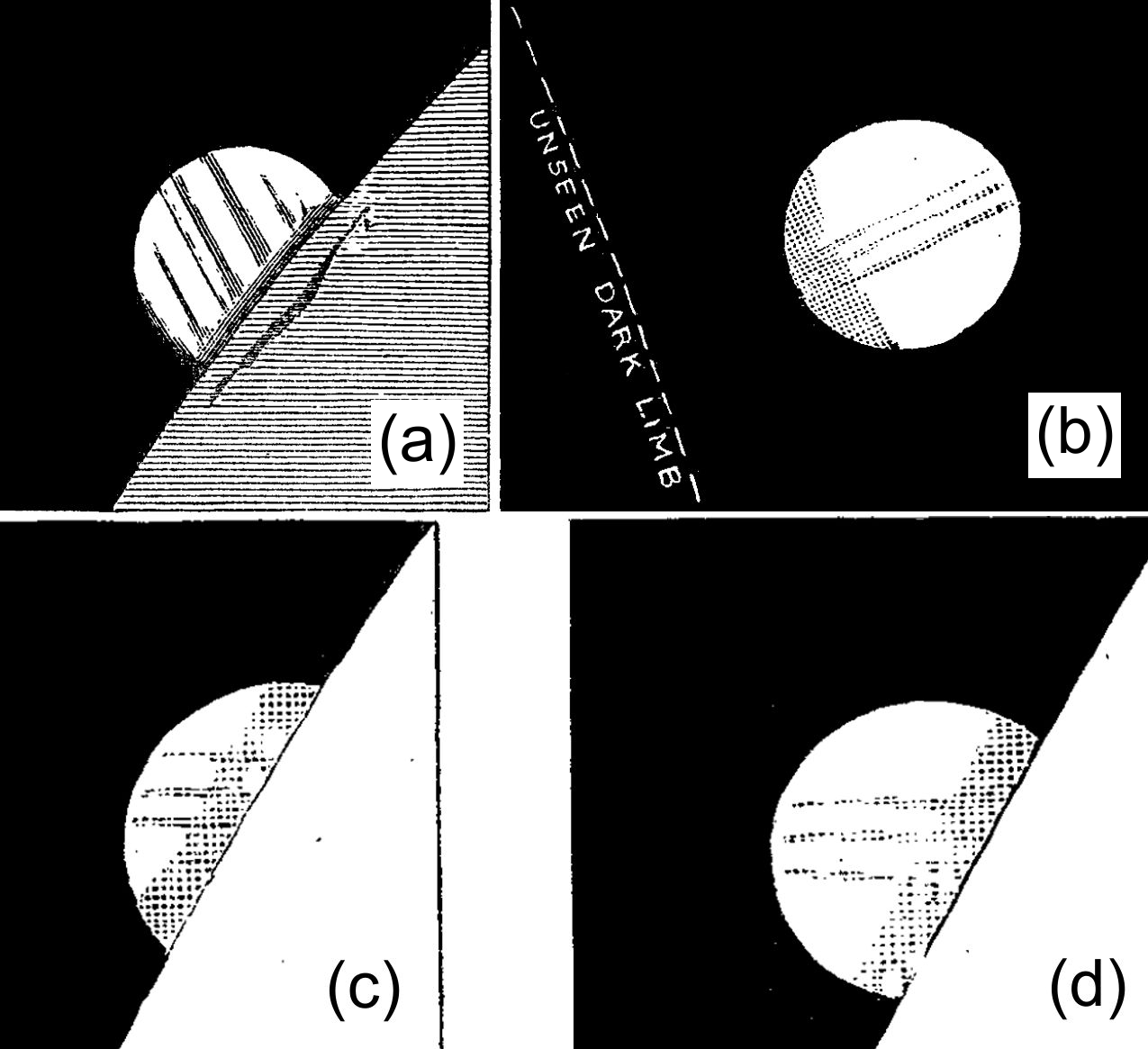}
   \caption{Examples of old reports on the dark band, seen above the lunar limb in the Jupiter background: \textbf{(a)} the jovian occultation on January 2, 1857 (Simms 1857);  \textbf{(b)} the immersion of the Jupiter on August 7, 1889, as it was seen by W. Wickham at Radcliffe Observatory in Oxford, using 3.25 inch  Marlborough telescope (Stone 1889); \textbf{(c)} and \textbf{(d)} the Jupiter appearance (emergence) on August 7, 1889, as it was seen by W.H. Robinson at the same observatory but using the 10-inch Barclay Equatorial (Stone 1889).}
              \label{Fig02}
    \end{figure}

During the next jovian occultation on January 2, 1857, W. Simms in London noted "\textit{the very positive line by which the Moon's limb was marked upon the planet; dark as the mark of a black-lead pencil close to the limb, and gradually softened off as the distance increased}" (Simms 1857). In Figure~\ref{Fig02}a we reproduce the corresponding sketch, provided by W. Simms in his report. This phenomenon was also confirmed by J. Simms, using another telescope. Independently from the above referred observers, professor J. Challis from the Cambridge Observatory saw \textit{"a faintish shadow running along the Moon's limb, and of about 4 arcseconds in breadth}" (Challis 1857). He specified: "\textit{The most remarkable phenomenon was the dark fringe bordering the Moon's limb: it was darker than Jupiter's belts; and continuing to the end of the emersion, made the observation of last contact somewhat difficult}".

The Jupiter occultation on August 7, 1889, was carefully observed in the Radcliffe Observatory in Oxford (Stone 1889). In that respect, W. Wickham reported on the occasion of the jovian immersion: "\textit{a shadow, like a crape veil, on the disc of Jupiter was noticed darkening that part of the planet nearest the Moon}" (see in Fig. ~\ref{Fig02}b). The analogous shadow was also seen during the emergence of Jupiter on the opposite side of lunar limb. The same darkening just before the Jupiter immersion was observed by W.H. Robinson with another telescope: "\textit{one quarter of the disc of the planet nearest the Moon's advancing limb being ill-defined and shaded, while the remaining 3/4 of disc was sharply defined}". A similar shadow was noted by him during the following up emergence of the planet (see in Fig. ~\ref{Fig02}c, d). Also Mr. F.A. Bellamy noted "\textit{no sharp edge}" of the Jupiter disc at first contact (immersion). At the same time, E.J. Stone did not see any shadow. Altogether, about 75\% of the observers of the Jupiter occultation on August 7, 1889 likely saw the phenomenon of dust extinction in conjunction with this event.

\begin{figure}
   \centering
   \includegraphics[width=0.375\textwidth,clip=]{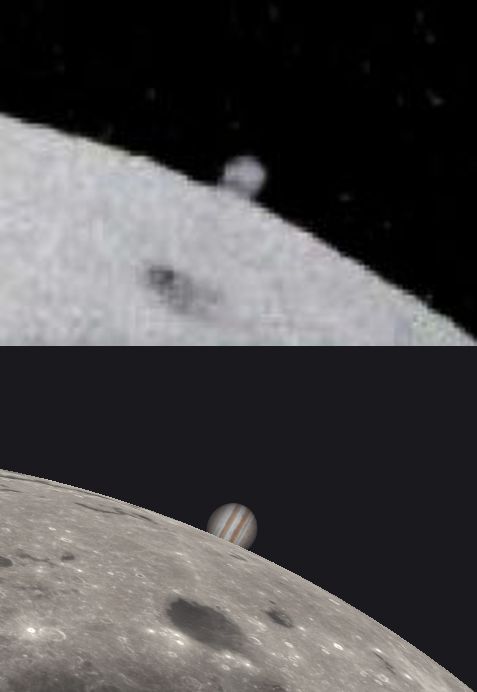}
   \caption{Photography (upper panel) of the dark band stretching along the lunar limb in the projection on Jupiter during its immersion on January 13, 1944 (Leavens 1944). Bottom panel shows the event simulation with the \textit{Stellarium} software.}
              \label{Fig03}
    \end{figure}

\begin{table*}
      \caption[]{Reported cases of the dust extinction effect near the lunar limb during planetary occultations.}
         \label{table01}
$$
            \begin{tabular}{c c c c c c c c c c}
            \hline
            \noalign{\smallskip}
            $\textrm{Date}^{(\textrm{a})}$ & UT$^{(\textrm{a})}$ & Pla- & Event & Limb$^{(\textrm{d})}$ & $H^{(\textrm{e})}$ & $W^{(\textrm{f})}$ & Observer & Long./lat. & Reference \\
            yyyy/mm/dd & hh:mm & net$^{(\textrm{b})}$ & type$^{(\textrm{c})}$ &  & km &  &  & deg. &  \\
            \noalign{\smallskip}
            \hline
            \noalign{\smallskip}
     1762/06/17 &  02:23 & S & E & D & 32$\pm$10 & 77.1$^{(\textrm{g})}$ & Dunn \ S. & -0.01/51.48 & Dunn \ 1762 \\
     1856/11/09 &  01:18 & J & G & NT & $-$ & 0 & Grove \ W.R. & -0.33/51.47 & Grove \ 1856 \\
     1856/11/09 &  01:18 & J & G & NT & $-$ & 0 & Goddard \ J.F. & -0.33/51.47 & Goddard \ 1856 \\
     1857/01/02 &  18:00 & J & E & B & 4.9$\pm$0.4 & 32 & Simms \ J. \& W. & -0.13/51.51 & Simms \ 1857 \\
     1857/01/02 &  18:00 & J & E & B & 7.1$\pm$0.1 & 32 & Challis \ J. & -0.13/51.51 & Challis \ 1857 \\
     1889/08/07 &  20:05 & J & I & D & 66$\pm$3 & 40 & Wickham \ W. & -1.26/51.76 & Stone 1889 \\
      &   &  &  &  &  &  & Robinson \ W.H. &  &  \\
     1889/08/07 &  21:00 & J & E & B & 32$\pm$19 & 40 & Wickham \ W. & -1.26/51.76 & Stone 1889 \\
      &   &  &  &  &  &  & Robinson \ W.H. &  &  \\
     1892/08/13 &  00:36 & J & I & B & 5$\pm$1 & 210 & Pickering \ W.H. & -71.53/-16.40 & Pickering 1892 \\
      &   &  &  &  &  &  & Douglass \ A.E. &  & Fitzpatrick 1944 \\
     1944/01/13 &  07:01 & J & I & B & 28$\pm$10 & 0 & Leavens \ P.A. & -73.09/40.75 & Leavens 1944 \\
      &   &  &  &  &  &  & White \ E.K. & -115.99/49.67 & Fitzpatrick 1944 \\
      &   &  &  &  &  &  & Morgan \ F.P. et al. & -73.66/45.50 & Wates 1944 \\
     1944/04/30 & 15:52 & J & E & B & 8$\pm$2 & 0 & Johnston \ H.M. & -77.61/43.18 & Fitzpatrick 1944 \\
      &   &  &  &  &  &  & Haas \ W.  & -75.29/39.95 & Haas 1944 \\
     1948/02/23 & 17:29 & M & E & B & 4 & 101 & White \ E.K. & -115.99/49.67 & Haas 1948 \\
      & 20:11 &  &  &  &  &  & Liston \ M. & -73.93/42.81 &  \\

                           \noalign{\smallskip}
            \hline
            \end{tabular}
         $$
     \textbf{(a)} Corrected for the modern time counting and verified with the virtual planetarium software \textit{Stellarium 1.2};
     \textbf{(b)} Notations: J (Jupiter), M (Mars), S (Saturn);
     \textbf{(c)} Notations: I (immersion), E (emergence), G (grazing);
     \textbf{(d)} Notations: B (bright limb), D (dark limb), NT (northern terminator);
     \textbf{(e)} The altitude $H$ of the upper border of dust cloud, estimated using the angular thickness of the dust band on planetary disk;
     \textbf{(f)} Daily total sunspot number (https://www.sidc.be/silso/DATA/SN\_d\_tot\_V2.0.txt);
     \textbf{(g)} Smoothed monthly mean sunspot number (https://ngdc.noaa.gov/stp/space-weather/solar-data/solar-indices/sunspot-numbers/depricated/international/tables/table\_international-sunspot-numbers\_monthly.txt)
    \label{table01}
    \end{table*}

Professor W.H. Pickering have published the photo of "\textit{a dark band three seconds in breadth seen stretching across the face of the planet,
tangent to the Moon's limb}" taken by him during the Jupiter occultation on August 13, 1892, at the Boyden Station in Arequipa, Peru (Pickering 1892).
His assistant, A.E. Douglass saw this dark band visually. However, the following up analysis of the photographs revealed that "\textit{although the dark band can be "faintly" seen by transmitted light on the negative, it does not show up in the prints}" (Fitzpatrick 1944).

Later on, professor W.H. Haas (Haas 1944) noticed a similar dark band in a photograph of Jupiter's occultation taken on January 13, 1944 (Fig. ~\ref{Fig03}). Independently, this band was observed by J.W. Campbell, F.P. Morgan, C.G. Wates and E.K. White  (Fitzpatrick 1944; Wates 1944).

In the years of 1944-1948 the occultations' dark band appeared in the scope of a special interest of the \textit{American Association of Lunar and Planetary Observers} (ALPO). This topic was regularly addressed in the ALPO proceedings "\textit{The Strolling Astronomer}", now known as the \textit{Journal of ALPO} (JALPO). However, it was considered mainly in the context of a possible lunar atmosphere. After the recognition of insignificance of the lunar gaseous envelope, the discussion of the occultation dark band stopped. Although H.C. Stubbs reported to ALPO on a possibility of clouds of the levitating lunar dust and their possible relation with the occultation anomalies (Haas 1955), the subject of the occultation dark band did not receive further attention since the middle of 1950s. Only a decade later, the existence of the lunar dust clouds was confirmed by in-situ measurements.

Nevertheless, the report on "\textit{a fine deep red line}" seen along the lunar dark limb successively on the background of the B-ring and the body of Saturn during its occultation on March 2, 1974, was included in the NASA's "\textit{Lunar transient phenomena catalog}" (Cameron 1978). The red color of the line in this case argues for the light extinction by a fine ($d \ll \lambda$) dust near the lunar surface.

\subsubsection{Long-lasting stellar occultations}

Besides of planetary light, the lunar dust clouds could extinct the stellar light too. The first known observer of such phenomena was F.D.J. Arago in the Royal Observatory of Paris. In particular, he reported that the star 49 Lib started to diminish in magnitude three or four seconds previous to its immersion (Grant 1852), while the normal duration of the immersing phase is of about of a few hundredths of a second.

Sometimes such anomalies were not confirmed by the observers in other locations at the distances of dozens kilometers (Tab. 2). For example, according to the reports of at least 4 observers at West Gorton in Manchester (Dawes 1863) and an independent person in Uckfield to the south of London (Leeson 1863), the immersion of $\kappa$ Cnc on April 26, 1863, continued for about of 0.5 s. However, the instantaneous disappearance of the same star was seen in London (Burr 1863) and in Nottingham (Vertu 1863). The long-lasting immersion ($\tau = 3.2$ s) of $\lambda$ Gem on March 6, 1884, was observed in Brighton, but it was not confirmed at 25 km from it, in Maresfield. In view of that, the long-lasting stellar occultations were often supposed to be caused by the terrestrial clouds.

\begin{figure}
   \centering
   \includegraphics[width=0.4\textwidth,clip=]{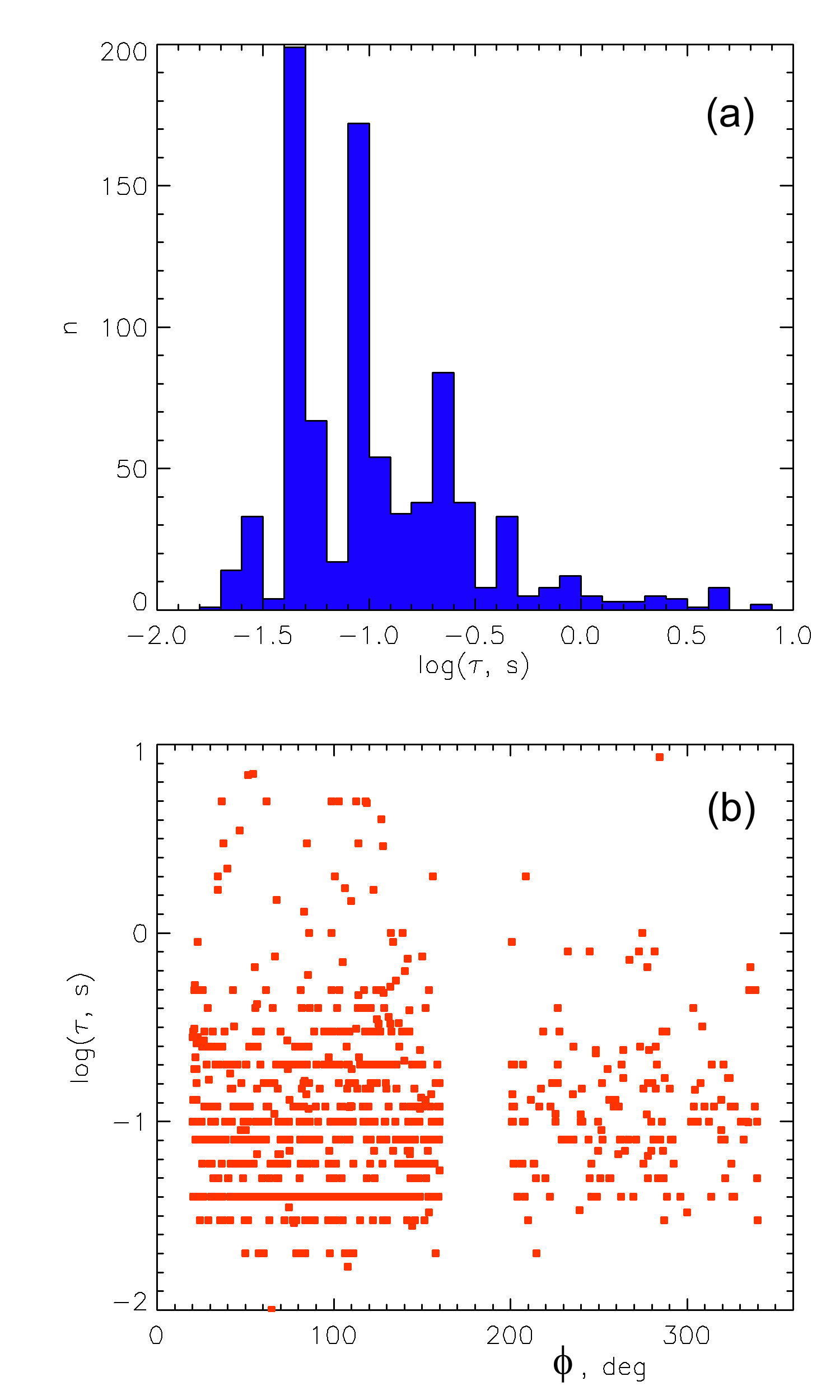}
   \caption{Duration time $\tau$ of the stellar occultations, extracted from the \textit{Lunar Occultation Archive} (Herald et al. 2022): \textbf{(a)} the total histogram of $\log (\tau)$; \textbf{(b)} the distribution of $\log (\tau)$ estimates versus the occultation axis angle $\phi$, counted
   eastwards in the terrestrial sky from the north pole of the Moon around the lunar limb. The grazing events near the poles with $0^{\textrm{o}}<\phi<20^{\textrm{o}}$, $160^{\textrm{o}}<\phi<200^{\textrm{o}}$, and $340^{\textrm{o}}<\phi<360^{\textrm{o}}$ are excluded.}
              \label{Fig04}
    \end{figure}

Using the new and extended data set of stellar occultations, we test the above mentioned hypothesis of the terrestrial cloudiness.
For this purpose we take the most complete data source, the so-called \textit{Lunar Occultation Archive} (thereafter LOA), provided in Herald et al. (2022)\footnote{\url{https://cdsarc.cds.unistra.fr/viz-bin/cat/VI/132C}}. It includes the historical data, as well as the collection of more recent observations enabled by the \textit{International Occultation Timing Association} (IOTA). Figure ~\ref{Fig04}a shows the histogram of all 2618 occultations over the years of 1967-2022 with the available estimates of their individual duration, $\tau$, extracted from LOA. Note, that the duration of each occultation event, controlled by both the Fresnel diffraction effects, and the effects of stellar diameter, usually should not exceed the values of several hundredths of a second. In the case of close double star occultations, the duration of event does not include to the time between the occultations of each companion. Nevertheless, the histogram in Fig.~\ref{Fig04}a has a pronounced extension at $\tau > 0.1$ s, demonstrating that 31\% (811) of all measured stellar occultations have anomalously high durations $\tau$. Figure~\ref{Fig04}b shows that such anomalies cannot be attributed with just grazing occultations. Even after exclusion of the grazing events, with the lunar limb crossing latitudes close to the poles, and leaving only those, for which the axis angle $\phi$ measured in the terrestrial sky eastwards from the Moon's north pole satisfies the conditions $20^{\textrm{o}}<\phi<160^{\textrm{o}}$ and $200^{\textrm{o}}<\phi<340^{\textrm{o}}$, there remain 418 long-lasting occultations with $0.1<\tau\leq8.6$ s. These include also the occultations in the equatorial zone with $\phi \sim 90^{\textrm{o}}$ and $\phi \sim 270^{\textrm{o}}$. It is worth to note here, that due to the specifics of the Moon motion in the sky, the visible trajectories of the occultating stars follow approximately the selenographic west-east direction, with the immersion and emergence events taking place at the selenographic western (eastern in the terrestrial sky) and eastern (western in the terrestrial sky) lunar limbs, respectively.

\begin{table*}
      \caption[]{Contradictory reports on anomalous occultations of stars.}
         \label{table02}
$$
            \begin{tabular}{c c c c c c c c c c}
            \hline
            \noalign{\smallskip}
            $\textrm{Date}^{(\textrm{a})}$ & UT$^{(\textrm{a})}$ & Star & Event & Limb$^{(\textrm{c})}$ & $\tau^{(\textrm{d})}$ & $W^{(\textrm{e})}$ & Observer & Long./lat. & Reference \\
            yyyy/mm/dd & hh:mm &  & type$^{(\textrm{b})}$ &  & s &  &  & deg. &  \\
            \noalign{\smallskip}
            \hline
            \noalign{\smallskip}
     1863/04/26 & 21:38 & $\kappa$ Cnc & I & D & 0.5 & 30 & Copeland \ R. & -2.20/53.47 & Dawes \ 1863 \\
           &  &  &  &  & 0.5 &  & Cooke \ T. \& sons & -2.20/53.47 &  -"- \\
           &  &  &  &  & 0.5 &  & Leeson \ C. & 0.10/50.97 & Leeson 1863 \\
           &  &  &  &  & $<0.1$ &  & Burr \ T.W. & -0.10/51.55 & Burr 1863 \\
           &  &  &  &  & $<0.1$ &  & Noble \ W. & 0.08/51.00 & Noble 1863 \\
           &  &  &  &  & $<0.1$ &  & Vertu \ J. & -1.17/52.97 & Vertu 1863 \\
     1884/03/06 & 22:09 & $\lambda$ Gem & I & D & 3.2 & 147 & Pratt \ H. & -0.15/50.82 & Pratt \ 1884 \\
           &  &  &  &  & $<0.1$ &  & Noble \ W. & 0.08/51.00 & Noble 1884 \\
                           \noalign{\smallskip}
            \hline
            \end{tabular}
         $$
     \textbf{(a)} Corrected for the modern time counting and verified with the virtual planetarium software \textit{Stellarium 1.2};
     \textbf{(b)} Notations: I (immersion), E (emergence), G (grazing);
     \textbf{(c)} Notations: B (bright limb), D (dark limb), NT (northern terminator);
     \textbf{(d)} Duration $\tau$ of the occultation;
     \textbf{(e)} Daily total sunspot number (https://www.sidc.be/silso/DATA/SN\_d\_tot\_V2.0.txt).
    \label{table02}
    \end{table*}

    \begin{figure}
   \centering
   \includegraphics[width=0.4\textwidth,clip=]{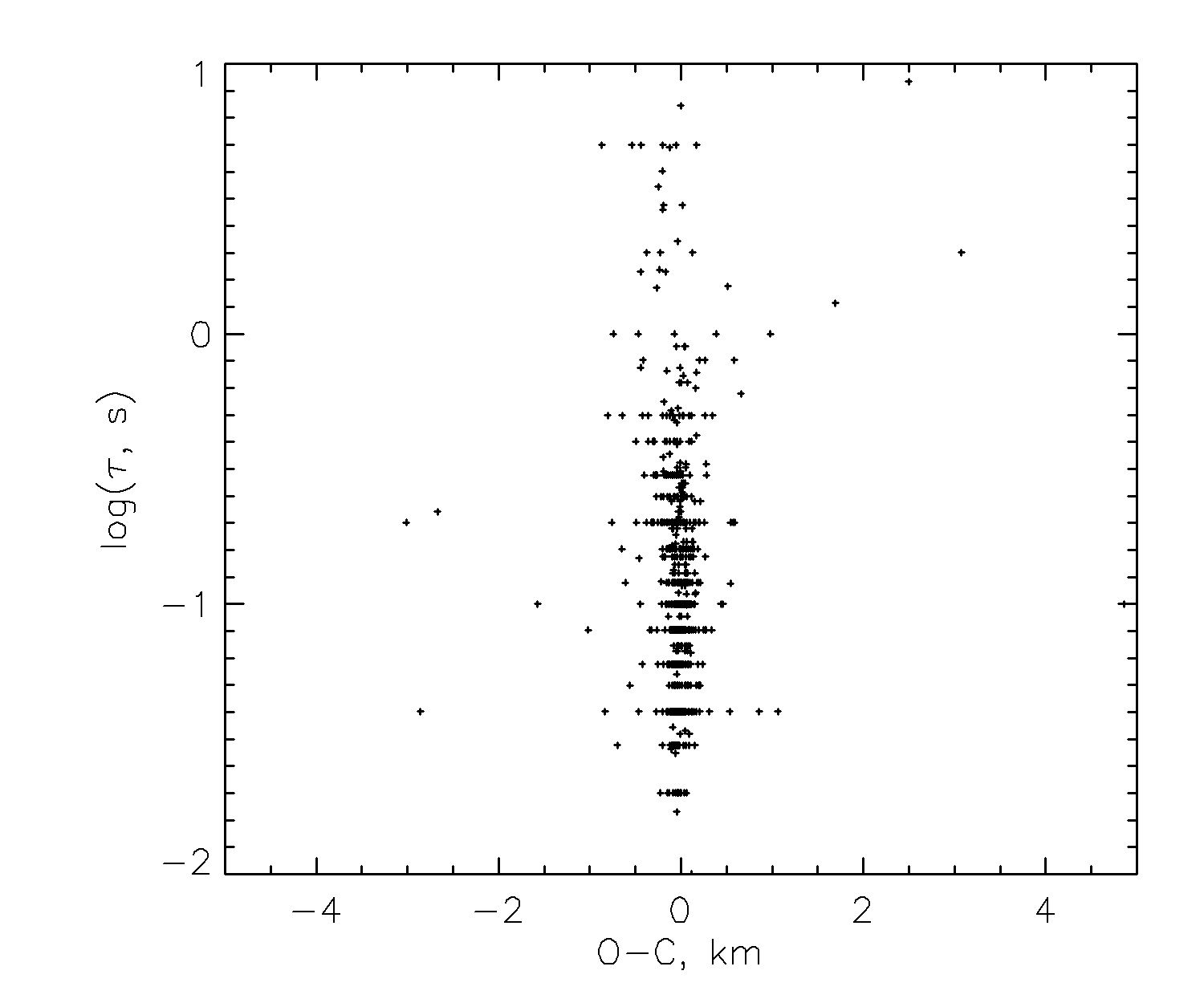}
   \caption{Duration time $\tau$ of stellar occultations versus the residual $O-C$ between the observed occultation altitude and the model lunar limb. The value $\tau$ as well as the corresponding $O-C$ are taken from LOA.}
              \label{Fig05}
    \end{figure}
    
In the case when not the lunar limb, but a terrestrial cloud occultates a star, the related pseudo-occultations happen at rather different heights above the lunar limb, being equally possible for any value of the occultation axis angle $\phi$. The hypothesis of terrestrial cloudiness can be verified by consideration of the provided in the LOA residuals, $O-C$, between the observed altitudes of the occultations and the corresponding calculated (modeled) levels of the lunar limb. In the case of significant contribution of the terrestrial clouds, the positive $O-C$ residuals should dominate. However, Fig.~\ref{Fig05} shows the clustering of the available $O-C$ residuals around zero for all events, including also the long-lasting ($\tau > 0.1$ s) occultations, showing no signs of an asymmetry towards positive $O-C$ residual values. Note, that the long-lasting events constitute a small part (0.16\%) of the whole content of LOA (including the occultations with unmeasured $\tau$), hence, they should not affect the model of the lunar limb, used for the estimation of $O-C$ residuals there. Moreover, the observed limitation of the long-lasting occultations at $\tau \leq 8.6$ s looks as a strange restriction for the hypothetic terrestrial clouds, but it seems to be a natural consequence of a decaying with height circumlunar dust concentration. \textit{Therefore, we conclude that the influence of terrestrial clouds is not dominant in the long-lasting occultation statistics.} Such non-terrestrial anomalies are a good argument for the starlight extinction in dust clouds above lunar limb.

\subsubsection{Shadows on an eclipsed Sun}

A light-absorbing dust cloud above the lunar surface should result in a shadow, observed on the background of the solar disc during solar eclipses. Such a phenomenon was observed by W.S. Jacob during the annular eclipse of the Sun on October 9, 1847. When the annulus was about being formed, the Moon's limb seemed to be united with that of the Sun by a dark ligament of about one minute of arc in breadth (Jacob 1848). The lunar limb was perfectly well defined, except at the part where the ligament was perceived. No such shadow was seen in course of the dissolution of the opposite side of the solar ring during the same eclipse.

During another solar eclipse on July 18, 1860, an expedition of the French Academy of Sciences observed in Algeria, and even photographed, a significant truncation or blunting of the horns of crescent sun (Laussedat et al. 1860). The lunar mountains were immediately dismissed as an explanation for the too strong distortion of the solar crescent: "\textit{It therefore only remains to invoke the action of a gaseous envelope around the Moon, an idea which spontaneously presents itself to the minds of observers}".

The similar horn truncation was noted during the solar eclipse on May 17, 1882, by professor Porschian with several witnesses in Istanbul (Flammarion 1882). For the same eclipse event, "\textit{a violent fringe, which made its appearance along the Moon's limb}" in the Sun (Hopkins 1882), as well as short arcs of the lunar limb visible outside the Sun (Noble 1882, Flammarion 1890) were reported.

\subsection{Light scattering phenomena}

In addition to the aforementioned light absorption effects, the circumlunar dust clouds are able to scatter light. The Mie theory of scattering predicts that, in the case of dust particles having diameters $d$ comparable with the light wavelength $\lambda$, the most effective scattering goes into the range of large phase angles $\theta \sim 180^{\textrm{o}}$, counted from the direction towards the light source. This is a so-called forward scattering. However, for the dust particles with $d \ll \lambda$, a backward scattering into the range of small phase angles $\theta \sim 0^{\textrm{o}}$ becomes as efficient as the forward scattering. In the case of macro-particles ($d \gg \lambda$), the dust, as well as the Moon itself, reflect the light, so that the maximum of reflected flux goes towards the small angles $\theta \approx 0^{\textrm{o}}$, which is known as an opposition effect. This effect might be useful for the search of large size dust particle ($d \gg \lambda$) formations, whereas forward scattering could indicate the presence of fine dust ($d \lesssim \lambda$) fraction in the lunar clouds.

\subsubsection{Forward scattering of sunlight}

The observation of forward scattering supposes large phase angles $\theta \approx 180^{\textrm{o}}$. At the same time, the sum of angles in the triangle Sun-Moon-observer is $\theta + \beta + \delta = 180^{\textrm{o}}$, where the angle $\beta$ is an angle between the observer-Moon and observer-Sun directions,  i.e., an angular distance between the Moon and the Sun in the observer's sky, whereas $\delta$ is an angular distance between the observer and Moon as seen from the Sun. The latter is a rather small value $\delta \lesssim 0.15^{\textrm{o}}$, which can be neglected, resulting in $\theta \approx 180^{\textrm{o}} - \beta$. Therefore, the above mentioned condition of the forward scattering observation requires a small $\beta$, which corresponds to the case of a crescent Moon. The same condition is also realized during solar eclipses by the Moon.

Apparently, the forward scattering of sunlight above the lunar surface was discovered by J.J. Schroeter (1792). On February 24, 1792, at the time interval from 17:05 to 18:45 UT, he observed the lunar "\textit{twilight}" in the form of "\textit{the very faint pyramidal glimmering light}", resembling the Zodiacal light and prolonging both lunar cusps along the dark limb of the new lunar crescent (about 2.5 days after solar conjunction). According to J.J. Schroeter's micrometric measurements, the greatest breadth of the twilight was only 2 arcseconds, while its length was 80 arcseconds, covering the arc of the limb of 4.89 degrees. The observer attributed this effect to a hypothetic lunar atmosphere and estimated the vertical height of the "\textit{inferior and more dense part of the Moon's atmosphere}" to be of about 0.4 km. This estimate appeared unrealistically low for an atmosphere, and was therefore disregarded. At the same time, such thicknesses are not forbidden for the dust clouds, which might range from several meters (Mishra \& Bhardwaj 2019) up to $\sim 100$ km (McCoy 1976). Therefore, the rejected long ago observational report needs reconsideration in the context of other data that may confirm its relation to the lunar dust.

In fact, the effect of transient prolongation of the lunar cusps, discovered by J.J. Schroeter, has a long history, which lasts from the 18th century up to modern time (Table 3). For example, Franz von Paula Gruithuisen, a professor of astronomy at the University of Munich, confirmed the J.J. Schroeter's observation, but at the date of April 16, 1838: "\textit{I saw the twilight clearly on both horns, most clearly on the northern horn, on either side between the horn and a very far-off mountain, which (i.e., twilight) lights up more dimly and extends somewhat beyond this mountain}." (Klein 1879). However, according to the Moon calender, no horns could be observed at this date on the Moon, which had the age of 20.5 days. At the same time, the well-defined horns of a young Moon crescent (age of 3.3 days) were seen a year later in the evening on April 16, 1839. Moreover, quite indicative is that F.P. Gruithuisen began his systematic search for the lunar twilights only in 1840, i.e., two years after his, mentioned above, first observation of the horns prolongation in 1838. Therefore, it is highly likely that the actual date of the lunar twilight observation reported by F.P. Gruithuisen was April 16, 1839, whereas the addressed contradiction with the Moon calender is due to the erroneously taken observation year of 1838, caused by false reading of the number "9" as "8" in the handwritten notebook of F.P. Grutheusen's (corrected in Table 3).

There exist also a number of documented observational anomalies during solar eclipses, which might be associated with the phenomenon of sunlight forward scattering by the lunar dust (Table 3). In particular, the Algiers Observatory, as well as several observers in Paris region, reported with regard of the partial solar eclipse on June 17, 1890, that "\textit{the lunar disc could be seen fairly well extending beyond the Sun at 3 or 4 minutes of arc from the edge}". According to the observer Ch. Trepied, "\textit{This extension of the lunar disk beyond the Sun can also be seen in a number of the photographs}" (Flammarion 1890). Moreover, similar phenomena (Figs. 6b,c) were noted in the reports on solar eclipses in 1875 and 1882 too (Noble 1875, 1882; Flammarion 1890).

    \begin{figure}
   \centering
   \includegraphics[width=0.3\textwidth,clip=]{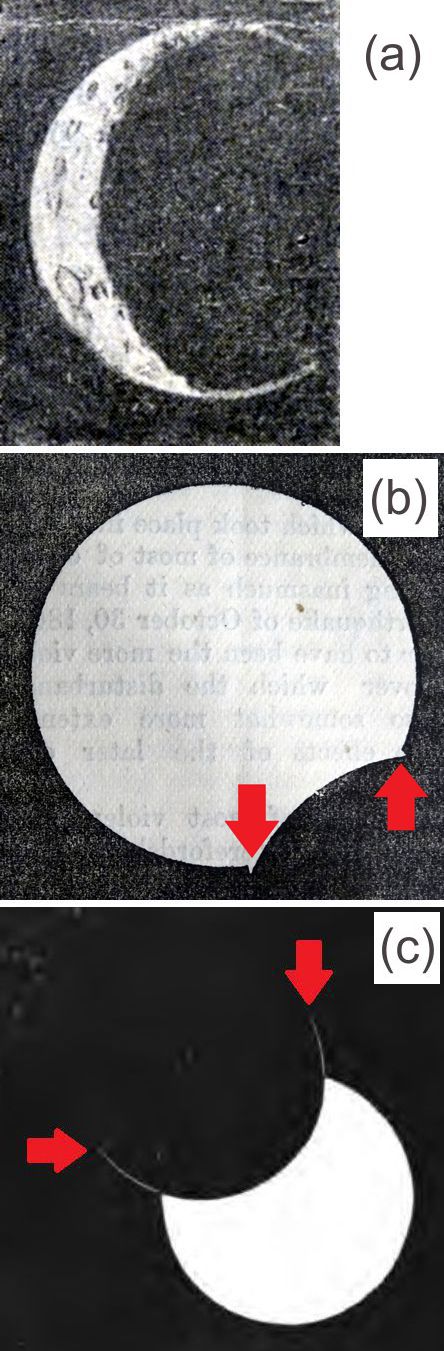}
   \caption{Lunar horizon glow seen from the Earth: \textbf{(a)} the prolongations of the lunar crescent horns on June 17, 1912 (Stolyarenko 1912);
   \textbf{(b)} the short ledges near the cusps of the solar crescent during the eclipse on September 29, 1875 (Noble 1875);
   \textbf{(c)} the long ledges near the cusps of the solar crescent during the eclipse on June 17, 1890 (Flammarion 1890).}
              \label{Fig06}
    \end{figure}

At least a dozen of reports on either prolongations of the crescent horns of the young Moon, or bright ledges near the cusps of the solar crescent during eclipses were found in literature published since 1877 (Fig.~\ref{Fig06}; Table 3). Some of them were included in the \textit{Lunar transient phenomena catalog} by NASA (Cameron 1978). However, a number of reports were disregarded as misinterpretations of the illuminated lunar mountains (bright points at a cusp), or because of the absence of estimations of their sizes, e.g., the extensions of the crescent horns. These extensions are measured as an angle $\gamma$, counted along the lunar limb from the normal cusp position to the end of the prolongation feature. They appear as light lines of 2-3 arcsecond thickness (Haas 1950a,b) of a bluish color (Arsyukhin 1994; case No. 285 in Cameron 1978; Florenskj \& Chernov 1994). The old reports on the lunar crescent horns' prolongations were confirmed by the modern Earth-based CCD/spectral/polarizational detections of the sunlight scattering by dust a cloud at terminator (Potter \& Zook 1992). There is an obvious analogy between the described prolongations and the glow of the lunar post-set horizon, photographed by the \emph{Surveyor} mission landers on the Moon (Rennilson \& Criswell 1974).

\begin{table*}
      \caption[]{Reports on the Moon crescent prolongations and ledges near the cusps of the eclipsed Sun crescent.}
         \label{table03}
$$
            \begin{tabular}{c c c c c c c c c c}
            \hline
            \noalign{\smallskip}
            $\textrm{Date}^{(\textrm{a})}$ & UT$^{(\textrm{a})}$ & $\beta^{(\textrm{b})}$ & Cusp & $\gamma^{(\textrm{c})}$ & $H^{(\textrm{d})}$ & $W^{(\textrm{e})}$ & Observer & Long./lat. & Reference \\
            yyyy/mm/dd & hh:mm &  deg. &  & deg. & km &  &  & deg. &  \\
            \noalign{\smallskip}
            \hline
            \noalign{\smallskip}
     1792/02/24 & $\sim$18:00 & 31 & N/S & 4.89 & $<$0.5 & 64$^{(\textrm{f})}$ & Schroeter \ J.J. & 8.91/53.15 & Schroeter \ 1792 \\
     1839/04/16 & $\sim$20:00 & 39 & N/S & - & - & 108 & Gruithuisen \ F.P. & 11.58/48.13 & Klein \ 1879 \\
     1840/02/08 & $\sim$19:00 & 64 & N/S & - & - & 168 & Gruithuisen \ F.P. & 11.58/48.13 & Klein \ 1879 \\
     1840/03/05 & $\sim$18:30 & 20 & N/S & - & - & 112 & Gruithuisen \ F.P. & 11.58/48.13 & Klein \ 1879 \\
     1840/03/06 & $\sim$19:00 & 33 & N/S & - & - & 115 & Gruithuisen \ F.P. & 11.58/48.13 & Klein \ 1879 \\
     1840/04/04 & $\sim$19:30 & $\approx$29 & N/S & 1 & $\lesssim$0.02 & 165 & Gruithuisen \ F.P. & 11.58/48.13 & Klein \ 1879 \\
     1840/06/01 & $\sim$19:30 & $\approx$21 & N/S & 2 & $\lesssim$0.04 & 82 & Gruithuisen \ F.P. & 11.58/48.13 & Klein \ 1879 \\
     1843/04/03 & 20:30 & 31 & N/S & 3.0 & $<$0.2 & 0 & Gruithuisen \ F.P. & 11.58/48.13 & Klein \ 1879 \\
     1875/09/29 & $\approx$14:10 & 0.46 & S/N & 0.08 & $\ll$0.001 & 192 & Noble \ W. & 0.08/51.00 & Noble \ 1875 \\
     1877/03/17 & 18:45 & $\approx$30 & S & - & - & 0 & Denett \ F. & -/- & Denett \ 1877 \\
     1882/05/17 & 09:33 & 0.42 & S/N & - & - & 192 & Noble \ W. & 0.08/51.00 & Noble \ 1882 \\
     -"- & $\approx$08:24 & - & S/N & - & - & 192 & Trepied \ Ch. & 3.03/36.80 & Flammarion \ 1890 \\
     1890/06/17 & 09:33 & 0.42 & S/N & - & - & 0 & Schmoll & 2.34/48.86 & Flammarion \ 1890 \\
     -"- & $\approx$11:14 & 0.29 & S/N & $\approx$30 & $\lesssim$0.001 & 0 & Herselin & 2.40/49.33 & Flammarion \ 1890 \\
     -"- & $\approx$11:05 & 0.14 & S/N & 13 & $\ll$0.001 & 0 & Trepied \ Ch. & 3.03/36.80 & Flammarion \ 1890 \\
     1912/06/17 & $\sim$20:00 & $\approx$35 & S/N & 40 & $\lesssim$35 & 12 & Stolyarenko \ D. & 31/48 & Stolyarenko 1912 \\
     1935/08/26 & $\sim$04:30 & $\approx$31 & NE & 60 & $\lesssim$49 & 48 & Wilkins \ P.H. & -0.13/51.51 & Wilkins 1951 \\
     1940/10/29 & 14:30 & 19 & N & 15 & $<$1.6 & 77 & Vaughn & -93.61/41.57 & Haas 1942 \\
     -"- &  & 19 & S & 5-10 & $<$0.7 &  &  &  &  \\
     1940/12/25 & $\sim$13:00 & $\approx$45 & N & 10 & $\lesssim$4.5 & 85 & Haas \ W.H. & -80.62/40.85 & Haas 1942 \\
     1950/06/19 & 03:35 & 39 & N & 12 & $<$4.7 & 110 & Haas \ W.H. & -106.65/35.12 & Haas 1950a \\
      -"- &  & 39 & S & 6 & $<$1.2 &  &  &  &  \\
     1950/07/20 & 03:22 & 59 & N/S & 12 & $<$12 & 184 & Haas \ W.H. & -106.65/35.12 & Haas 1950b \\
     1956/03/14 & 19:00 & 26 & S & 21 & $<$5.9 & 173 & Firsoff \ V.A. & -2.86/51.09 & Cameron 1978 \\
     1982/11/19 & 14:20 & 43 & N & 4.5 & $<$0.8 & 158 & Filonenko \ V.S. & 36.23/50.00 & Florenskj \& Chernov 1994 \\
     1983/01/19 & 16:05 & 59 & S & 2.7 & $<$0.6 & 127 & Filonenko \ V.S. & 36.23/50.00 & Florenskj \& Chernov 1994 \\
                            \noalign{\smallskip}
            \hline
            \end{tabular}
         $$
     \textbf{(a)} Corrected for the modern time counting and verified with the virtual planetarium software \textit{Stellarium 1.2};
     \textbf{(b)} Angular distance between the Moon and the Sun in the sky, according to \textit{Stellarium 1.2};
     \textbf{(c)} The size of observed prolongations/ledges as an angle $\gamma$, counted along the lunar limb from the normal cusp position to the end of feature;
     \textbf{(d)} The altitude of scattering dust, according to Eq. (\ref{Eq2});
     \textbf{(e)} Daily total sunspot number (https://www.sidc.be/silso/DATA/SN\_d\_tot\_V2.0.txt);
     \textbf{(f)} Smoothed monthly mean sunspot number (https://ngdc.noaa.gov/stp/space-weather/solar-data/solar-indices/sunspot-numbers/depricated/international/tables/table\_international-sunspot-numbers\_monthly.txt)
    \label{table03}
    \end{table*}

    \begin{figure}
   \centering
   \includegraphics[width=0.4\textwidth,clip=]{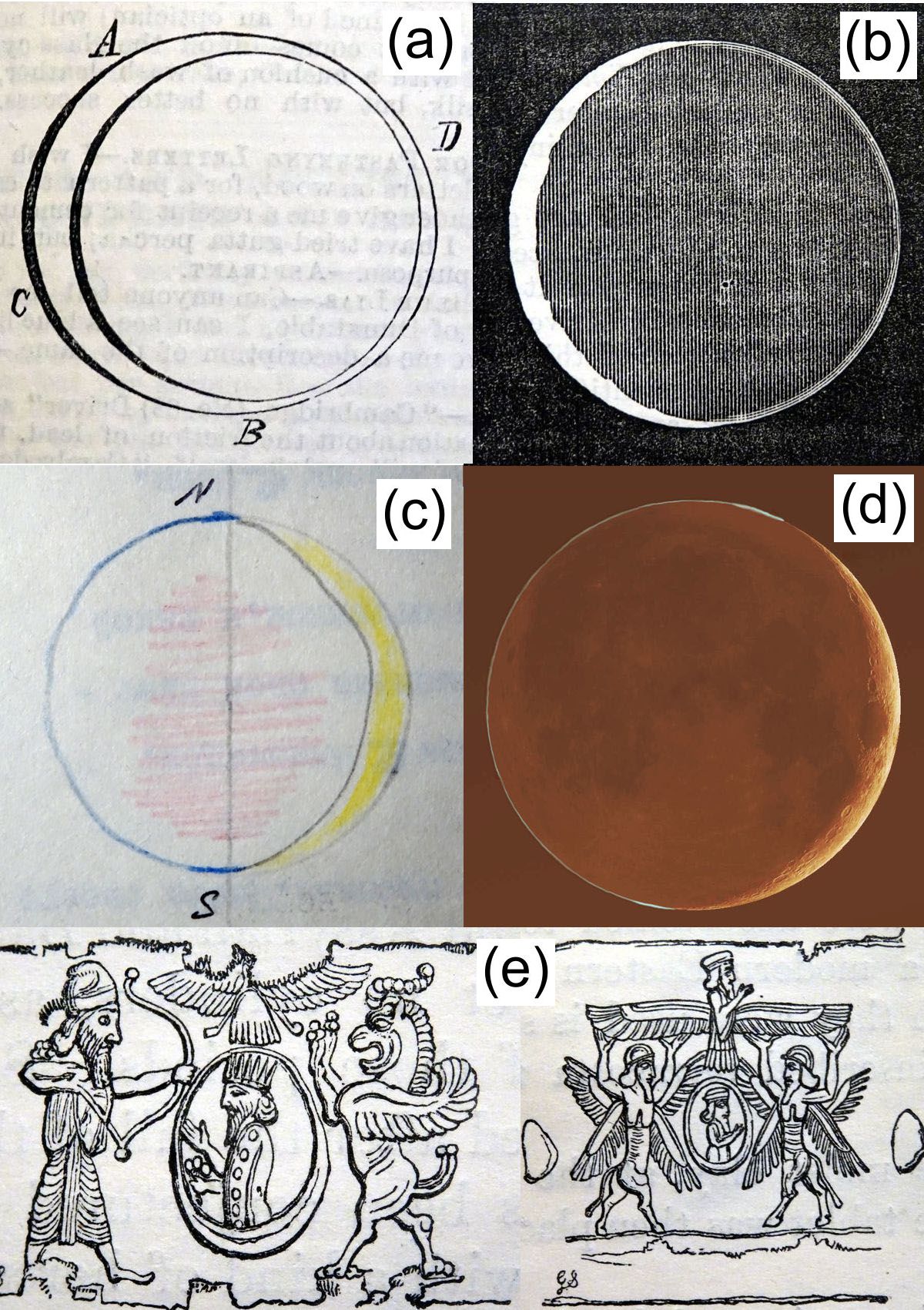}
   \caption{Annular Moon images: \textbf{(a)} The scheme of the phenomenon seen on January 03, 1867 (Bickerdike 1867);
   \textbf{(b)} The gravure of the Moon, published in 1882 (Ranyard 1882);
   \textbf{(c)} The observer's hand drawn sketch of the phenomenon seen by E. Arsyukhin on December 18, 1982 (personal communication)
   \textbf{(d)} The simulation of the event drawn in (c), using \textit{Stellarium 1.2} software;
   \textbf{(e)} The ancient Persian images from Babylon (Layard 1853).}
              \label{Fig07}
    \end{figure}
    
   In the extreme case, the lunar horns prolongations with $\gamma \geq 90^{\textrm{o}}$ close the crescent, and turn it into a ring. We call this, so far unnamed, special observational phenomenon as an "\emph{annular Moon}". At the same time, it is worth to say that the images of annular Moon (i.e., a thin glowing contour around the dark lunar limb from one cusp to another forming the lunar annulus) can be found in ancient artifacts and pieces of art, like e.g., in the Persian cylinders (Fig.~\ref{Fig07}e) from Babylon (Layard 1853). Apparently, the first reported observation of an annular Moon is dated by January 3, 1867, when "\textit{in addition to the eastern crescent, there appeared a well defined illuminated annular space}" (Bickerdijk, 1867). Unfortunately, a precise location of the observer of this event is unknown, whereas the provided details (the Moon altitude of about $26^{\textrm{o}}$ and "a planet a little to the south") point to the region of Bengaluru in India.

There were attempts to proclaim the annular Moon as an illusion of contrast effect or an increased albedo of the lunar highlands at the limb.
However, the narrow glowing ring around the lunar dark side was confirmed using the telescopes (Corliss 1985), and it was even repeatedly photographed (Hoffmann 2017). The observers have often noted that this ring of light differs from the ashen light of the Moon by its silvery or bluish color (e.g., Corliss 1985; Arsyukhin 1994). Consequently, such reports are listed in Table 4, as a special part of the whole set of considered dust phenomena.

In the cases when the time and/or location of an observer of the annular Moon was not reported, we recovered this information by modeling of the observation circumstances, using the virtual planetarium software \textit{Stellarium 1.2}. This approach enables substituting the unknown time of observation with its estimate, which corresponds to an approximately optimal observational conditions, as a compromise  between the lunar altitude above the horizon and the crepuscular illumination of the sky. The uncertainty of such estimate is of about 0.5-1 hour, which gives an acceptable error $\varepsilon \lesssim 0.5^{\textrm{o}}$ of the determined Sun-Moon angular distance $\beta$. For the majority of found angles $\beta > 20^{\textrm{o}}$ this error stays within the range of $\lesssim 2.5$\%. The observer location coordinates were estimated using the Google Map service, by taking the towns (i.e., geographic locations) named in the original reports or in other publications of the same authors. The performed survey and simulations made it possible to estimate finally the heights $H$ of the dust, illuminated by the Sun (see in Section 3).

\begin{table*}
      \caption[]{Reports on the annular Moon observations (i.e., when $\gamma \geq 90^{\textrm{o}}$).}
         \label{table04}
$$
            \begin{tabular}{c c c c c c c c}
            \hline
            \noalign{\smallskip}
            $\textrm{Date}^{(\textrm{a})}$ & UT$^{(\textrm{a})}$ & $\beta^{(\textrm{b})}$ & $H^{(\textrm{c})}$ & $W^{(\textrm{d})}$ & Observer & Long./lat. & Reference \\
            yyyy/mm/dd & hh:mm &  deg. & km &  &  & deg. &  \\
            \noalign{\smallskip}
            \hline
            \noalign{\smallskip}
      1867/01/03 & 06:30 & 32 & $>$70 & 0 & Bickerdike & 77.6/13.0 & Bickerdike 1867 \\
      1882/-/- & - & $\sim$30 & $\gtrsim$61 & - & Ranyard \ A.C. & -0.11/51.52 & Ranyard 1882 \\
      1882/-/- & - & $\sim$30 & $\gtrsim$61 & - & Webb \ T.W. & -0.11/51.52 & Ranyard 1882 \\
      1882/11/07 & 05:00 & 41 & $>$117 & 133 & Hopkins \ B.J. & -0.13/51.51 & Hopkins 1883 \\
      1883/03/12 & 20:00 & 50 & $>$180 & 95 & Hopkins \ B.J. & -0.13/51.51 & Hopkins 1883 \\
      1912/03/12 & $\approx$06:30 & $\approx$74 & $\gtrsim$438 & 27 & Henderson \ A.C. & $\approx$-0.13/51.51 & Corliss 1985 \\
      1912/11/07 & 06:37 & $\approx$22 & $\gtrsim$33 & 0 & Henderson \ A.C. & $\approx$-0.13/51.51 & Corliss 1985 \\
      1914/03/28 & $\sim$18:30 & $\approx$21 & $\gtrsim$30 & 0 & Henderson \ A.C. & $\approx$-0.13/51.51 & Corliss 1985 \\
      1915/01/17 & $\sim$17:00 & $\approx$26 & $\gtrsim$46 & 78 & Stevenson \ E.A. & $\approx$-0.13/51.51 & Corliss 1979 \\
      1915/02/16 & $\sim$18:00 & $\approx$30 & $\gtrsim$61 & 47 & Stevenson \ E.A. & $\approx$-0.13/51.51 & Corliss 1979 \\
      1916/01/07 & 17:55 & 35 & $>$84 & 127 & Stevenson \ E.A. & $\approx$-0.13/51.51 & Corliss 1979 \\
      1916/02/06 & 18:03 & 40 & $>$112 & 137 & Stevenson \ E.A. & $\approx$-0.13/51.51 & Corliss 1979 \\
      1916/12/26 & 16:33 & 25 & $>$42 & 68 & Stevenson \ E.A. & $\approx$-0.13/51.51 & Corliss 1979 \\
      1917/12/17 & 16:58 & 43 & $>$130 & 252 & Stevenson \ E.A. & $\approx$-0.13/51.51 & Corliss 1979 \\
      1918/03/14 & 19:02 & 27 & $>$49 & 177 & Stevenson \ E.A. & $\approx$-0.13/51.51 & Corliss 1979 \\
      1918/12/08 & $\sim$17:30 & $\approx$60 & $\gtrsim$269 & 113 & Pickston \ J. & $\approx$-0.13/51.51 & Corliss 1985 \\
      1950/06/19 & 03:35 & 39 & $>$106 & 110 & Haas \ W.H. & -106.65/35.12 & Haas 1950a \\
      1950/07/20 & 03:22 & 59 & $>$259 & 184 & Haas \ W.H. & -106.65/35.12 & Haas 1950b \\
      1957/07/31 & 02:24 & 54 & $>$213 & 212 & Craig \ J. & -97.33/37.70 & Haas 1957 \\
      1982/12/18 & $\sim$16:00 & $\approx$34 & $\gtrsim$79 & 136 & Arsyukhin \ E. & 37.62/55.75 & Arsyukhin 1994 \\
      2006/-/- & - & 17 & $>$19 & - & Hoffmann \ M. & 28.64/38.55 & Hoffmann 2017 \\
      2008/-/- & - & 19 & $>$24 & - & Hoffmann \ M. & 10.65/49.20 & Hoffmann 2017 \\
      2011/-/- & - & 11 & $>$8 & - & Hoffmann \ M. & 10.65/49.20 & Hoffmann 2017 \\
                            \noalign{\smallskip}
            \hline
            \end{tabular}
         $$
     \textbf{(a)} Corrected for the modern time counting and verified with the virtual planetarium software \textit{Stellarium 1.2};
     \textbf{(b)} Angular distance between the Moon and the Sun in the sky, according to \textit{Stellarium 1.2};
     \textbf{(c)} Altitude of the scattering dust, according to Eq. (\ref{Eq1});
     \textbf{(d)} Daily total sunspot number (https://www.sidc.be/silso/DATA/SN\_d\_tot\_V2.0.txt);
    \label{table04}
    \end{table*}
    
    \subsubsection{Forward scattering of planetary light}

The forward scattering effects could be observed not only for the sunlight, but also for the light, which comes from the planets.
Indeed, there exist several (quite rare) reports regarding observations of the light glows, seen at the edge of the lunar disk emanating from the occultated planets, still well behind the Moon's limb (see ALX12 section in Corliss 1985).

In particular, during the grazing occultation of Jupiter on October 19, 1968, "\textit{an arc of light over the (occultated) planet's position and just above the Moon's dark limb}" was observed (Brock 1969).
On October 17, 1973, about 1-2 s before the appearance of Saturn from behind the dark lunar limb, "\textit{a faint glow ... which appeared like seeing 'a campfire on the other side of a treeless hill on a dark night'}" was seen (Reed et al. 1974).

\subsubsection{Forward scattering of starlight}

The lunar dust clouds illuminated by a starlight could be also observed sometimes. Several reports on such events are listed below.

On April 22, 1913, a slight pre-cursor of $\pi$ Sco appeared more than one second before the star emergence above the daylight lunar limb  (Gheury 1913).  A cloud veil as a cause was ruled out due to the distinct appearance of the nearby Langrenus crater. The observer proclaimed that this phenomenon gives an "\textit{evidence of an atmosphere on the Moon}".

On July 24, 1915, when the star $\phi$ Sgr came close to the dark limb of the Moon, but a certain distance still remained between them, the star began to stretch into a strip, three times longer then its width, and then it disappeared instantly (Barabashov 1915; case No. 357 in Cameron 1978).

On June 6, 1933, about 90 s before the appearance of Regulus, a thin white and sharply bordered line was observed along the illuminated limb of the Moon, separated by a thin and very dark line right above the limb (Luizard 1933).

On February 25, 1972, a slight light of the star BD$+21^{\textrm{o}}$ 1724 was seen during 0.5 s after the occultation (Chernov \& Florenskj 1975).

\subsubsection{Backscattering of the sunlight}

Reflecting dust clouds could be especially detectable on the background of dark sky. Theoretically this effect means that the sunlight is scattered in small phase angles $\theta < 90^{\textrm{o}}$, i.e., the angular distance between the Moon and the Sun appears to be sufficiently large $\beta = 180 - \theta > 90^{\textrm{o}}$. While such a so-called opposition effect increases the brightness of the Moon, it could consist also in the sunlight reflection from large lunar-dust particles at $\theta \approx 0^{\textrm{o}}$. Therefore, the conditions of an almost full Moon are favorable for the search of circumlunar dust clouds of large grains.

\begin{figure}
   \centering
   \includegraphics[width=0.45\textwidth,clip=]{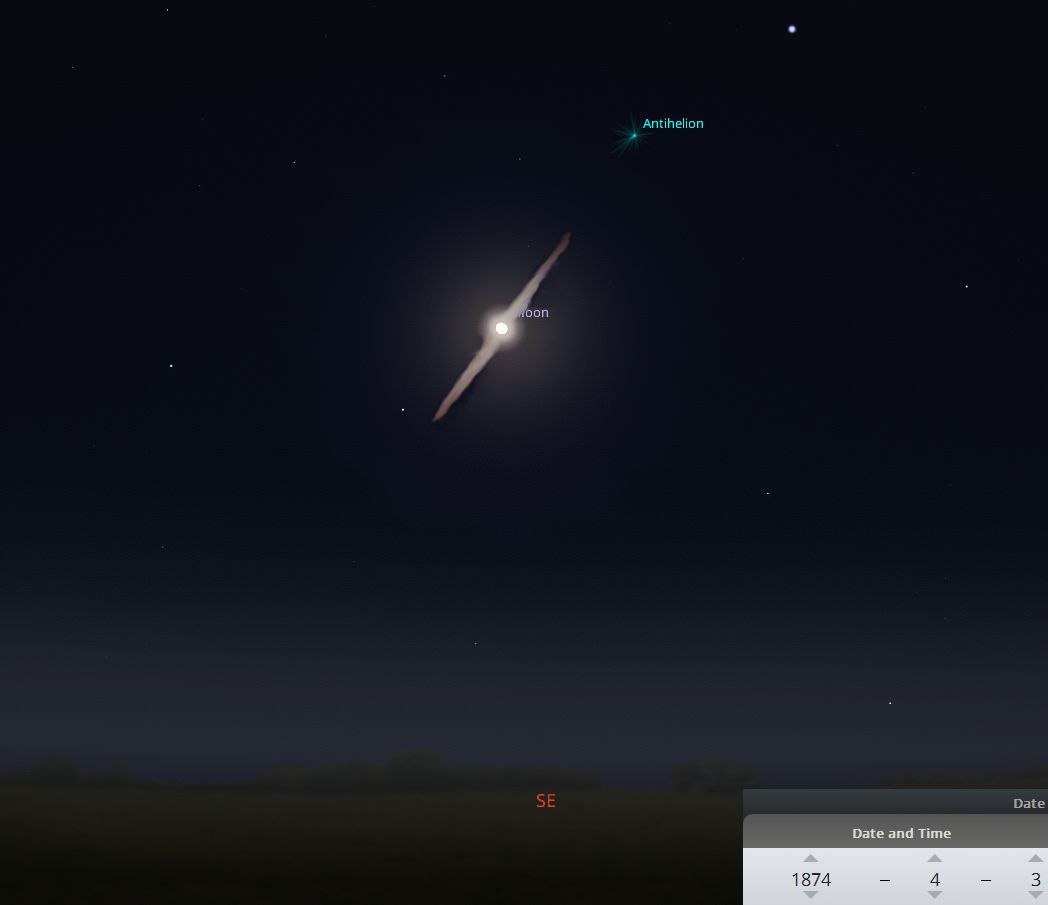}
   \caption{Reconstruction of the lunar zodiacal light seen on April 3, 1874, at 22:15 UT in Cambridge (Massachusetts) according to the report of observer; L. Trouvelot (Corliss 1985). The virtual planetarium software \textit{Stellarium 1.2} was used for the reconstruction. }
              \label{Fig08}
    \end{figure}

In this context, the old reports on strange luminous formations, associated with an approximately full Moon, are worth of attention. Among those are, for example, the testimonies on the phenomenon of \emph{lunar zodiacal light}, seen as luminous cones of light on either side of the Moon, extending approximately along its orbital plane (see, e.g., the reports collection in Corliss 1985). In particular, such phenomenon was noted on April 3, 1874 in in Cambridge (Massachusetts) by an authority of observational astronomy, L. Trouvelot (Fig.~\ref{Fig08}). Of a similar nature could be also the \emph{Lunar protuberances}, seen on July 13, 1875 (Loftus 1875) and on November 12, 1878 (Hammer \& Rodgers 1878).

Some cases of \emph{visibility of the terrestrial shadow edges} beyond the eclipsed Moon could be also attributed to manifestations of the lunar dust clouds (Florenski and Chernov 1973, 1975). However, several of such phenomena were proven to be related with the thin layer of terrestrial cloudiness (Lenard et al 1892, Flammarion 1880).

Numerous reports regarding the "lunar clouds", "mists", and "obscuration" (Cameron 1978; Corliss 1985), in fact, are connected with the natural manifestations of sufficiently dense dust clouds, observed in front of (i.e., projected) the lunar disk. In particular, the appearance and evolution of the local dust cloud in the crater Langrenus was instrumentally confirmed in the \textit{Observatoire de Paris} using CCD images, as well as polarimetry measurements (Dollfus 2000). There exist also other instrumental confirmations of the reality of such local dust clouds (see, e.g., in Cameron 1978).

\section{Phenomenology of dust clouds}

As follows from the review in previous section, during the last three centuries a lot of information has been accumulated regarding the observed manifestations of lunar dust clouds, accessible for remote sensing. The collected historical material allows the study of properties and statistics of the lunar dust cloudiness. However, the phenomenological difference between particular observational cases, demonstrates the complexity and variety of the exiting dust formations. Correspondingly, our analysis below is focused on different scales of the lunar dust cloud patterns.

\subsection{Large-scale cloud pattern}

We begin our analysis from the global phenomenon of an annular Moon. Its geometry allows a simple estimation of an altitude (thickness) $H$ of the circumlunar dusty envelope necessary to produce the light annulus around the dark limb of the lunar crescent.

\begin{figure}
   \centering
   \includegraphics[width=0.45\textwidth,clip=]{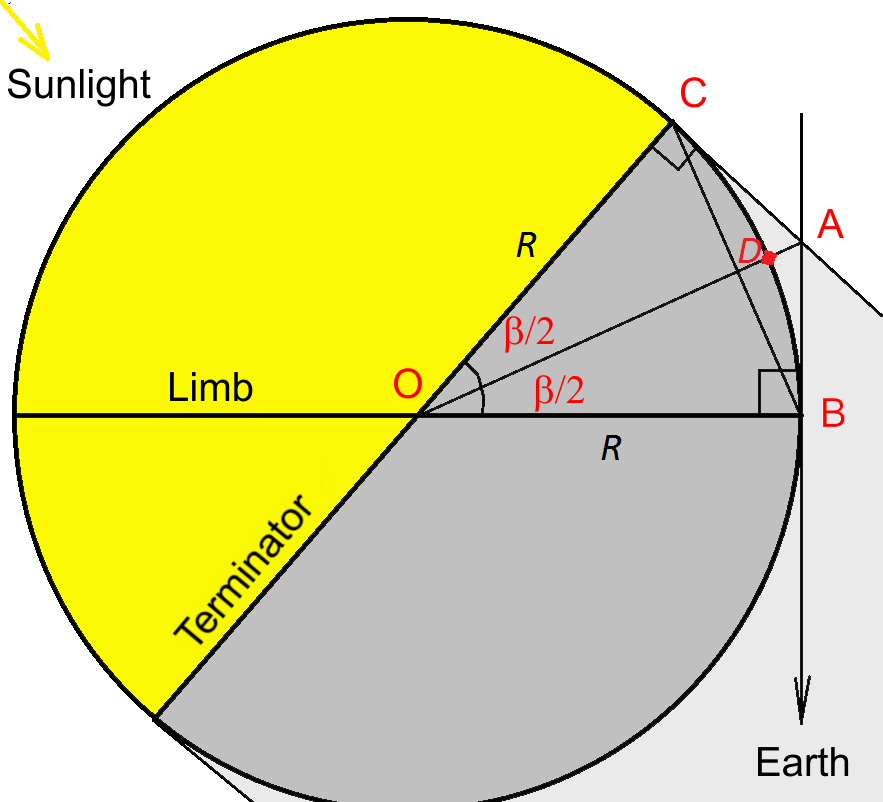}
   \caption{Lunar projection onto the Sun-Moon-observer plane. The altitude of the edge of the lunar shadow, i.e., the distance $DA$, as it is seen by an Earth-based observer along the line of sight $AB$, corresponds to the minimal height $H$ of the circumlunar dusty envelope, which scatters the sunlight and produces the effect of the annular Moon. The lunar radius $R$ and the visible Sun-Moon angular distance $\beta$ in the sky, allow finding the value of $H$.}
              \label{Fig09}
    \end{figure}

Figure~\ref{Fig09} shows the lunar projection onto the Sun-Moon-observer plane. The point O is the lunar center, which is projected onto lunar cusps, i.e., the intersections of the terminator and the limb line (i.e., of the visibility horizon).

Let us to find the altitude of the edge of the lunar shadow, i.e., the distance $DA$, as it is seen by an Earth-based observer along the line of sight $AB$. It corresponds to the minimal height $H$ of the sunlight scattering circumlunar dusty envelope, which produces the effect of the crescent horns prolongation with $\gamma = 90^{\textrm{o}}$, i.e., the formation of annular Moon. The lunar radius $R$ and the angle $\beta = \angle COB$, introduced in subsection 2.2.1 as the Sun-Moon angular distance in the sky, allow to find $H$, i.e., the distance $DA$, by considering the triangle $\triangle OAB$. In particular, given that the distance $OA$ is equal to $R/\cos(\beta/2)$, we obtain that the sunlight, scattered by the dust, located \emph{above}

\begin{equation}
 H = R(\sec(\beta/2) - 1),
 \label{Eq1}
\end{equation}
will be seen from the Earth above the dark lunar horizon (at the level of point $B$). Using the virtual planetarium software \textit{Stellarium 1.2} and Eq. (\ref{Eq1}), we calculated the values of $\beta$ and corresponding $H$ for all known cases of the observed annular Moon, listed in Table 4.

The estimates of $H$ in Table 4 show that the minimal altitudes for the sunlight scattering dust range from 8 km up to $H \gtrsim 438$ km. In about 40\% of all cases (8 of 23), the dust is localized at the altitudes $H>100$ km, which appear in the area of parking orbits of the space probes and maned missions. In the rest 60\% of the listed cases the altitudes of circumlunar dusty envelope cover the range of lower orbits, crucial for descending trajectories of the landers. It seems to be just a lucky coincidence that, the annular Moon phenomenon did not take place during the active phases of the previous lunar space missions.

The blue or silver (due to the reddening of the observed Moon's crescent near horizon) color of the light contour along the dark lunar limb (Arsyukhin 1994; Corliss 1985) argues in favor of the Rayleigh scattering mechanism of the sunlight by a finest ($d \ll \lambda$) fraction of the levitating dust, which has the scattering cross-section $\sigma_{s} \varpropto \lambda^{-4}$, decreasing with the wavelength increase.

It is difficult to estimate the timescale (i.e. duration) of the known annular Moon phenomena, because of their connection with the relatively short-lasting continuous visibility period of the lunar crescent (typically $\beta / 15 \ \textrm{deg/h} \lesssim 2$ hours) and the daily periodicity of consequent observations. Note that according to Tab.4, there are no reports on an annular Moon which was observed over several days. Altogether, this information limits the timescale of an annual Moon between $\sim 2$ and 24 hours.

\subsection{Short-scale cloud pattern}

An analog of the above considered large-scale dust clouds, resulting in an annular Moon phenomenon, exists also among the short-scale dust formations. In particular, the Moon crescent prolongations are a less grandiose phenomenon. Usually, they were observed in the circumpolar regions, where the sunlight illuminates the low-altitude part of the dust clouds. Their real spatial extent is, however hidden in the entire dark space. Nevertheless, the altitude $H$ of the visible part of such polar dust clouds can be estimated.

To find the value of $H$, instead of the plane projection in Fig.9, we consider a more general case, taking into account the sphericity of the Moon. In particular, the triangle $\triangle OAB$ in Fig.~\ref{Fig09} turns to a spherical one, where the arc $OB$ characterizes the crescent prolongation with an angle $\angle OAB = \gamma$. Using the known solution for a right spherical triangle, we express $\tan(\angle AOB) = \sin(\angle OAB) \tan(\beta/2) = \sin(\gamma) \tan(\beta/2)$. Correspondingly, $\sec(\angle AOB) = \sqrt{\tan(\angle AOB)^{2} + 1}$. A displacement along the line of sight, tangential to the surface of the Moon, from the point $B$ on the limb to the shadow edge at $A$ increases the altitude $H=R \sec(\angle AOB) - R$. The value $H$ exceeds the thickness of dust cloud beyond the crescent prolongation. Hence, the angle $\gamma$ of a maximal prolongation corresponds to the illuminated dust located \emph{below} the altitude
\begin{equation}
 H = R\left\{\sqrt{\left[\sin \gamma \tan\left(\frac{\beta}{2}\right)\right]^{2} + 1} - 1\right\},
 \label{Eq2}
\end{equation}

At $\gamma = 90^{\textrm{o}}$, the right hand side of Eq. (\ref{Eq2}) turns to that of Eq. (\ref{Eq1}). The same Eq. (\ref{Eq2}) can be applied also for the estimation of $H$ in the case of bright ledges near the cusps of solar crescent observed in a few cases during the solar eclipses (like in Fig. \ref{Fig06}b,c). In such cases the cusp points in Fig. \ref{Fig09} correspond to the position of projection of the solar limb onto the lunar one, and related geometry constructions for the derivation of Eq. (\ref{Eq2}) remain the same as in Fig.\ref{Fig09}. According to the obtained estimates of $H$, provided in Table 3, the dust clouds related with the crescent prolongation phenomenon, are located at relatively low altitudes of $H \lesssim 49$ km with a typical value of $H \backsim 1$ km.

Besides of the crescent prolongations and bright ledges, also the stellar occultations can be used for probing of the short-scale (close to surface) and the most dense parts of the dust clouds. In particular, the duration time $\tau$ of an occultation can be converted into a thickness of the dust cloud $L=\tau V_M$, where $V_M = 1.022$ km/s is an average orbital velocity of the Moon. The occultations are seen in different points of the lunar limb at various values of the occultation axis angle $\phi$, specified above in subsection 2.1.2. Observing numerous occultations, one can study the behavior of an average value of the dust cloud thickness $\langle L \rangle = \langle \tau \rangle V_M$ as a function of the angle of incidence $i \approx 90^{\textrm{o}} - \phi$, which is an angle between the visible trajectory of an occultated star and the normal direction to the lunar limb in the occultation point. Here we again take into account that all stellar occultation trajectories are approximately parallel to the selenographic west-east direction. In result, one could obtain a generalized profile of the dust cloud.

\begin{figure}
   \centering
   \includegraphics[width=0.40\textwidth,clip=]{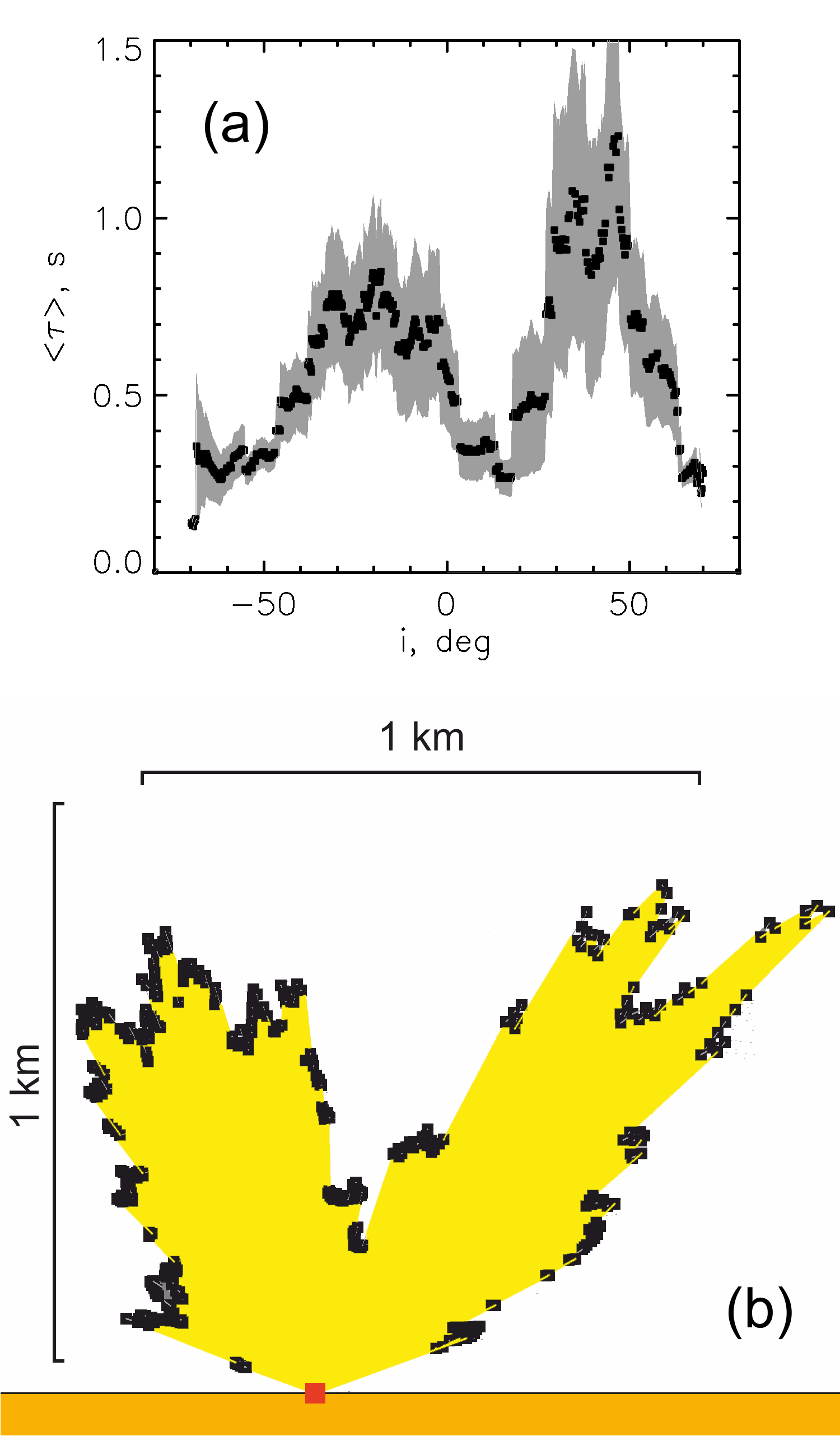}
   \caption{Average duration $\langle \tau \rangle$ of anomalous occultations with $\tau > 0.1$ s from the LOA (Herald et al. 2022) versus the angle of incidence $i$. \textbf{(a)}: the diagram in cartesian coordinates, where black squares show mean values of $\tau$   averaged over the sliding window of $\delta i =20^{\textrm{o}}$ width, and the corresponding standard-deviation range is filled with gray color. \textbf{(b):} the diagram in polar coordinates, which depicts the generalized profile of the short-scale dust cloud (filled with yellow color). The red square marks the place of star occultation by the lunar limb (orange color).}
    \label{Fig10}
    \end{figure}

Figure ~\ref{Fig10}a shows the distribution of the value $\langle \tau \rangle$, versus the incidence angle $i$, averaged over a sliding window of $\delta i = 20^{\textrm{o}}$ width. This diagram is constructed only for the anomalously long-lasting ($\tau > 0.1$ s) occultations, which are associated with the lunar dust clouds. One can see that $\langle \tau \rangle$, and therefore the average cloud thickness $\langle L \rangle$ are minimal at $i \approx \pm 70^{\textrm{o}}$, which correspond to the polar regions, according to the relation $\phi = 90^{\textrm{o}} - i$.

 It is worth noting here, that in the case of a global spherical circumlunar dusty envelope with an approximately constant thickness (i.e. altitude), in contrast to the measured fact (Fig. ~\ref{Fig10}a), one can expect for the stars occultated at the close to polar regions of the lunar limb ($\phi \sim 0$ and $i \sim \pm 90^{\textrm{o}}$) the longer paths behind the dust cloud, and therefore higher values of $\tau$, as compared to those for the stars occultated at the limb equator ($\phi \sim \pm 90^{\textrm{o}}$ and $i \sim 0$). Nevertheless, the diagram in Fig. ~\ref{Fig10}a shows quite a different pattern of $\langle \tau \rangle$ with the minima at $i \rightarrow \pm 90^{\textrm{o}}$, as well as at $i\approx 15^{o}$. This diagram, recalculated for $\langle L \rangle = \langle \tau \rangle V_M$ and presented in the polar coordinates in Fig. ~\ref{Fig10}b, shows a contour outline of a generalized dust cloud with two well defined lobes directed at the angles of $i \sim 45^{\textrm{o}}$. Therefore, the revealed geometry of dust cloud corresponds to the standard model view of an impact plume with a hollow cone and the material ejection angles within a limited range around $i = 45^{\textrm{o}}$ (see, e.g., in Cheng et al. 2020). Transillumination of such ejecta cone by a star results in two lobes of the maximal extinction, when the stellar light ray is tangent to the cone walls. This feature of an impact plume is well resembled in the dust cloud pattern in Fig. ~\ref{Fig10}b. Some asymmetry of the cloud could be interpreted as a result of drift of the levitating dust in the electric field of inhomogeneous surface charge. The average plume size, revealed from the typical values of the long-lasting occultation time, is of about 1 km. This fact explains the contradictory reports on the durations of some stellar occultations seen quasi-simultaneously, but at different places at tens of kilometers apart (see Table 2).

Altogether, our interpretation of the obtained results as a cumulative manifestation of the sporadic short-scale dust clouds near the surface, produced by the meteoroid impacts, is in accordance with the mainstream point of view. The latter considers the meteoroid impacts as a main source of the circumlunar dust clouds, according to the data obtained in-situ from the LADEE mission (Hor\'{a}nyi et al. 2015).

\subsection{Transitive dust formations}

It is worth to note, that the vertical scale of about 1 km of the plume-like dust formations (Fig.~\ref{Fig10}b) is close to the typical thickness $H \sim 1$ km of the dust clouds responsible for the prolongations of the lunar crescent horns (Table 3) and dark bands seen during planetary occultations (Table 1). This correspondence suggests the connection between these phenomena. Apparently, the plume-like compact dust-clouds could be source of more extended dust formations, including the quasi-global dusty envelopes discussed in Subsection 3.1.
Hence, the transitive dust formations could be present on the Moon as well. They may be connected with a specific evolution of the considered above large- and short-scale dust formations, or related with other processes on the lunar surface and in interior.

\begin{figure}
   \centering
   \includegraphics[width=0.40\textwidth,clip=]{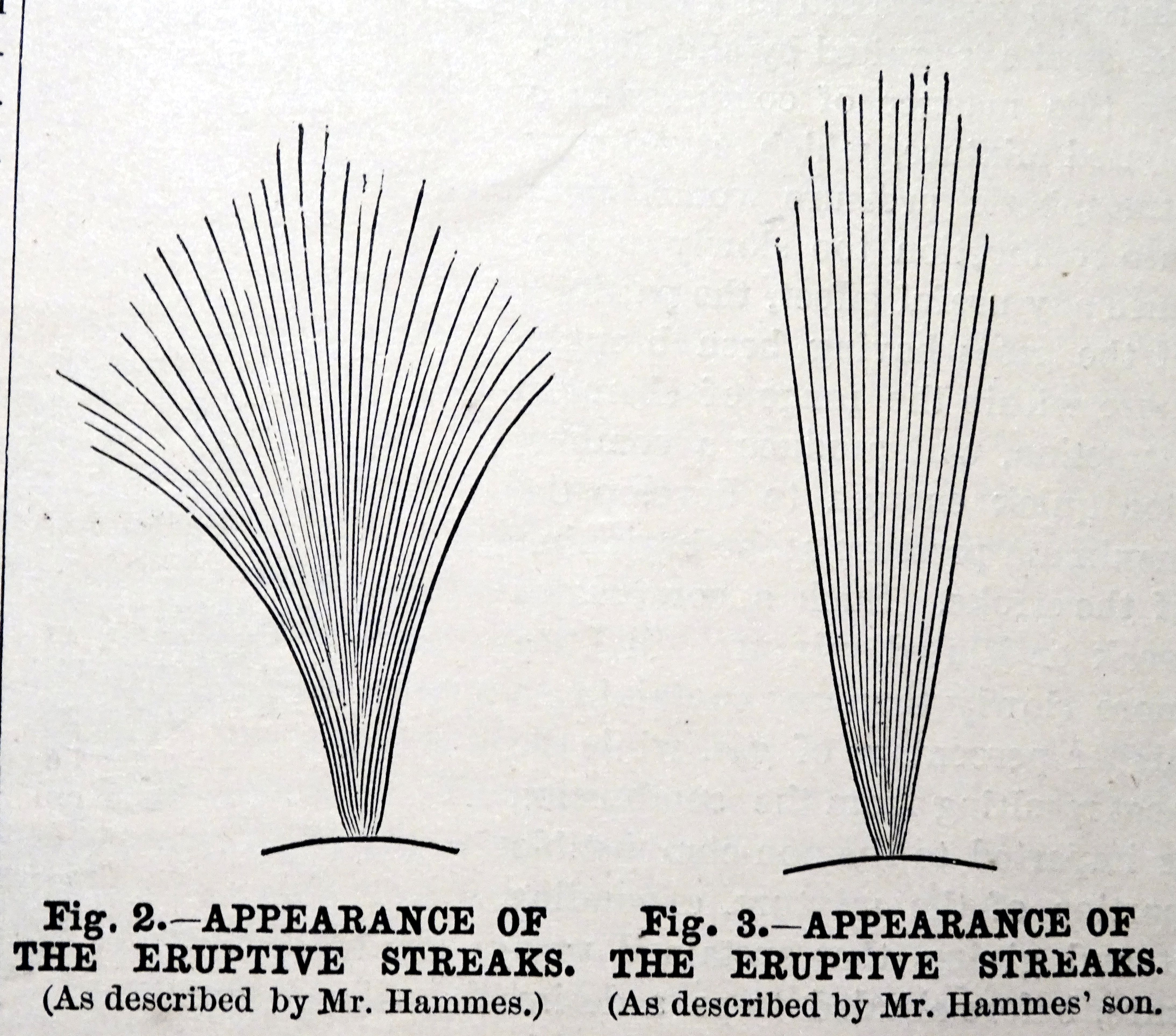}
   \caption{"\textit{"Supposed volcanic eruption on the Moon}" on November 13, 1878, as it was presented by the U.S. Naval Observatory (Rodgers \& Hammer 1878; No. 208 in Cameron 1978).}
              \label{Fig11}
    \end{figure}

Indeed, there are reports on lunar "protuberances" mentioned, e.g., in Subsection 2.2.4, which could be considered as an intermediate type of phenomena between the local and global dust clouds (Fig. ~\ref{Fig11}). Although the heights of such dust formations are comparable with altitude of the global circumlunar dust envelope, which results in the annular Moon (Table 4), their horizontal scales are significantly smaller and correspond to the scale of large craters.

In particular, a "light fountain" of 162 km height in the Aristarchus region at terminator has been photographed 4 times at the Passau Observatory on April 25, 1972 (K\"{u}veler \& Klemm R. 1972). Such dust fountains or "streamers", appeared as light bands illuminated by the Sun, were observed in situ by the astronauts of the Apollo 10 and 17, as well as possibly Apollo 8 and 15, missions. It was concluded: "\textit{The angular extent of the streamers ($>30^{\textrm{o}}$ fully developed) indicate that the light scattering particulates extended from the lunar surface to above the orbital altitude of the spacecraft}" (McCoy \& Criswell 1974). The local electrified dust fountains near twilight craters were found also in the data of the LADEE lunar orbiter mission (Lianghai et al. 2020).

\subsection{Appearance statistics}

The impact related nature of the most of dust clouds, detected during the stellar occultation events with anomalously long duration $\tau \geq 0.1$ s, can be verified using their appearance statistics. To decrease the influence of observational selection on this statistics, we consider the probability of anomalous occultation events $P = n/N$ in the LOA, where $n$ is the number of long-lasting occultations in a bin of histogram, and $N$ is the total number of occultations in the same bin.

\begin{figure}
   \centering
   \includegraphics[width=0.45\textwidth,clip=]{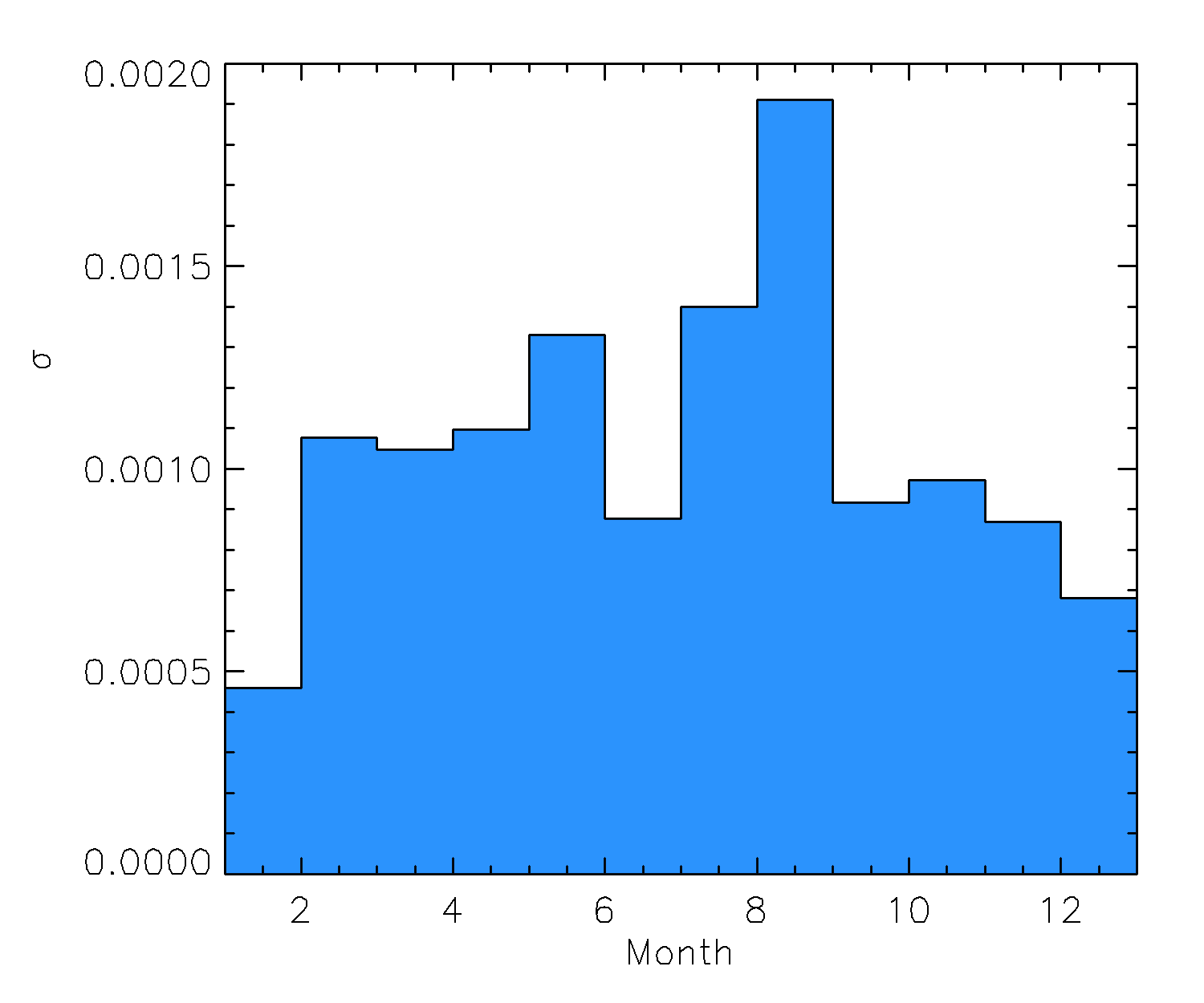}
   \caption{Probability $P$ of the long-lasting ($\tau \geq 0.1$ s) occultation events distributed over a year (i.e. versus  months), according to the LOA data.}
              \label{Fig12}
    \end{figure}

\begin{figure}
   \centering
   \includegraphics[width=0.40\textwidth,clip=]{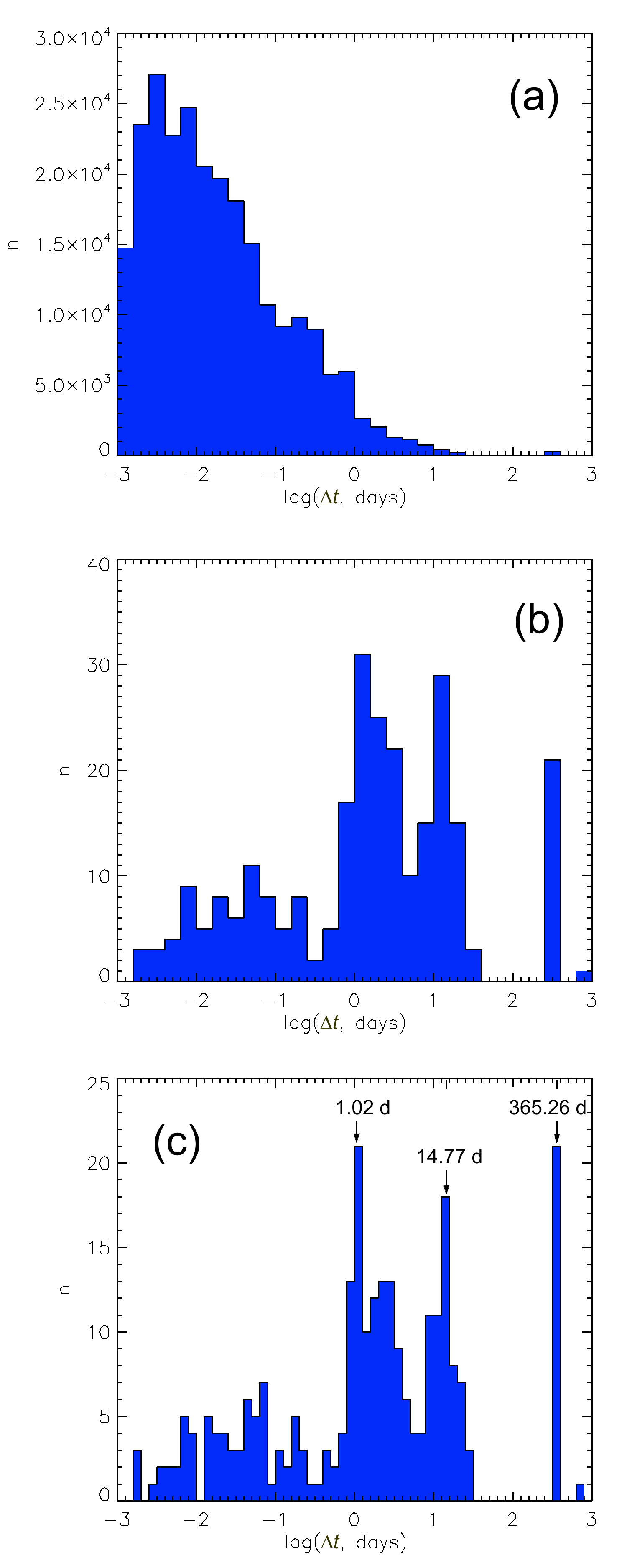}
   \caption{Histograms of time intervals $\Delta t$ between successive stellar occultation events available in LOA: \textbf{(a)} for all occultations, including also those with the measured $\tau$; \textbf{(b)} for the long-lasting occultations with $\tau \geq 0.1$ s only; \textbf{(c)} same as in panel (b), but with a higher resolution of the histogram. The annual periodicity (365.26 days), lunar culminations (1.03 day), and the half of synodic lunar month (14.77 days) are arrowed for comparison.}
              \label{Fig13}
    \end{figure}

Figure~\ref{Fig12} shows the distribution of probability $P$ of the long-lasting occultation events over a year (i.e., versus months). One can see that the maximal probability is in August, i.e., when the most abundant meteor shower of Perseids is active. The corresponding annual periodicity in the appearance of the long-lasting occultation events is also seen in the histograms in Fig.~\ref{Fig13} of the time intervals $\Delta t$ (from LOA) between the successive occultation events. Although, the annual peak is barely distinguishable in the distribution of $\Delta t$ constructed for all available occultation events in the LOA, (Fig.~\ref{Fig13}a), it becomes more prominent in the statistics of anomalous occultations with $\tau \geq 0.1$ s (Fig.~\ref{Fig13}b,c).

Besides of the annual periodicity, there is a clear peak at 14.77 days in the histogram in Fig.~\ref{Fig13}c, which corresponds to the half of lunar synodic month. It supposes the solar tidal influence with the same period. This half-month peak can not be due to the lunar phase (varying illuminated part of the Moon's body seen from the Earth), affecting the visibility of occultations. Otherwise, a noticeable pick should be present in the histogram at $\log(\Delta t)=1.47$, which corresponds to the main period of the lunar illumination (synodic month of 29.53 days). Such pick is, however, absent in Fig.~\ref{Fig13}b,c. The one-day peaks in diagrams in Fig.~\ref{Fig13}b,c are an effect of observational selection, related with the daily periodicity of occultations measurements, mainly during the night hours.

Altogether, the statistical annual periodicity of the anomalous occultations and their maximum at the season of high meteoroid showers confirm the hypothesis of an impact related nature of the majority of dust clouds around the Moon. At the same time, the presence of additional periodicities in the statistics of the long-lasting stellar occultations and related circumlunar dust formations indicate about additional factors which affect their appearance. We would like to note in that respect, that the local dust clouds, seen on the background of the lunar disk as the "Darkenings" and "Gaseous" type of the lunar transient phenomena (Cameron 1978), correlate with the anomalistic month (i.e., a period of the Moon passing through the perigee of its geocentric orbit every 27.554551 days). So, they are interpreted as a result of the terrestrial tidal impact on the Moon (Cameron 1977). However, due to the mostly non-limb location of such phenomena on the lunar disc, they cannot result in anomalous stellar occultations and produce a detectable effect in their statistics. This explains the absence of the corresponding monthly periodicity peaks in the diagrams on Fig.~\ref{Fig13}b,c. At the same time, the effect of solar tidal influence on the lunar transients remains unstudied. Nevertheless, its period of 14.77 days, well seen in Fig.~\ref{Fig13}b,c, confirms the role of the varying gravity (in particular, solar one) in the generation and modulation of the lunar dust clouds along with the meteoroid impacts.

A possibility of the lunar outgassing to lift up the dust has been often supposed and discussed (e.g., Crotts \& Hummels 2009 and therein). Anomalous stellar occultations, archived in the LOA, allow to study the distribution of outgassing sites on the surface of the Moon. Indeed, there is no way to distinguish between the manifestations of dust clouds produced by the lunar eruptive events and those generated by the meteoroid impacts, as they all equally result in the anomalous long-lasting stellar occultations. However, one may hope to see any specific features in the latitudinal distribution of the dust formations around the Moon, which could be associated with the potential source sites on the lunar surface.

\begin{figure}
   \centering
   \includegraphics[width=0.40\textwidth,clip=]{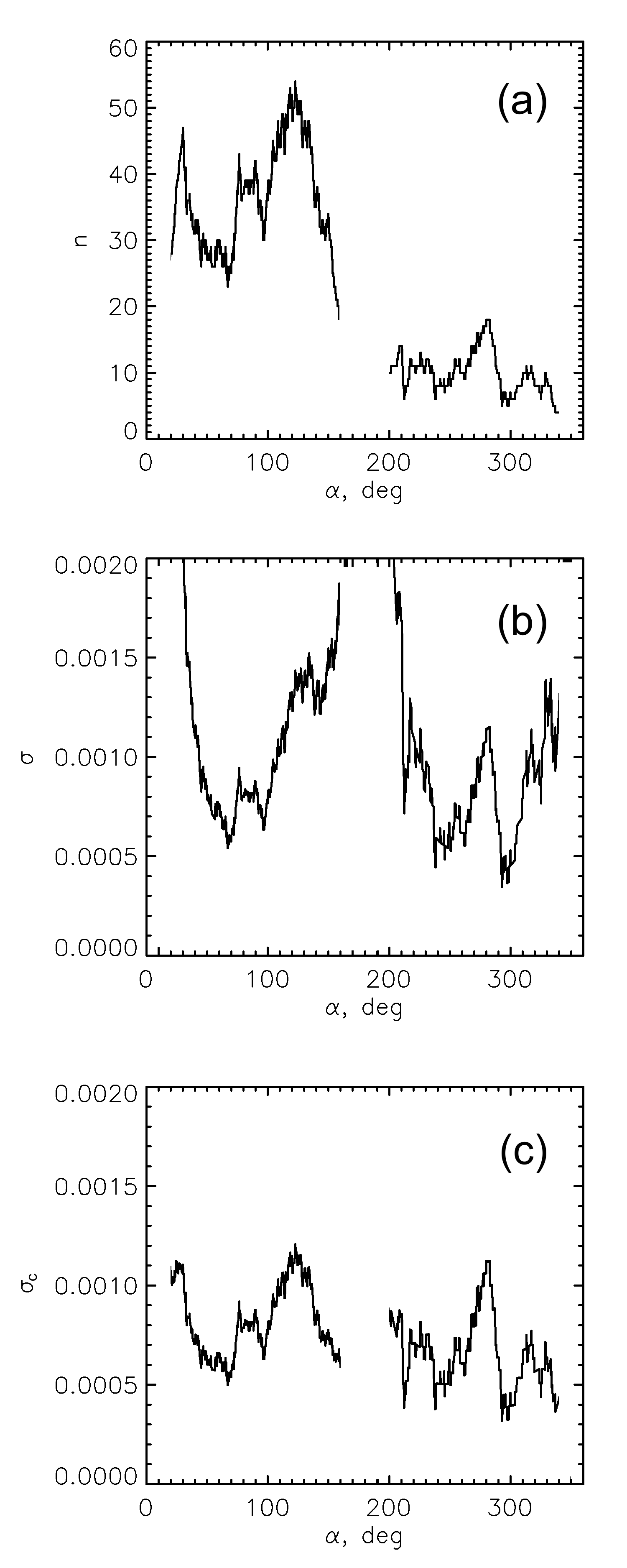}
   \caption{Distribution of the long-lasting stellar occultation events from the LOA with $\tau \geq 0.1$ s, averaged in a sliding window of $\delta \phi = 20^{\textrm{o}}$ width, versus the occultation axis angle $\phi$: \textbf{(a)} the number of long-lasting occultations $n$ in the sliding window; \textbf{(b)} the occurrence probability $P = n/N$, where $N$ is the number of all occultations in the sliding window (including also normal events with $\tau < 0.1$ s) ; \textbf{(c)} the corrected occurrence probability $P_{c} = n |\sin \phi|/N$. The occultations in $\pm 20^{\textrm{0}}$-regions around the poles are excluded to avoid of the grazing occultations.}
              \label{Fig14}
    \end{figure}

Figure~\ref{Fig14} shows the main features of the distribution of the long-lasting ($\tau \geq 0.1$ s) occultations versus the occultation axis angle $\phi$, averaged over a sliding window of $\delta \phi = 20^{\textrm{o}}$ width. In particular, according to Fig.~\ref{Fig14}a, anomalous occultations dominate in the eastern part of the limb (with $\phi < 180^{\textrm{o}}$) with the maximal amount taking place at $\phi \approx 120^{\textrm{o}}$. At the same time, the opposite side of the limb at $\phi > 180^{\textrm{o}}$ shows a deficit of the long-lasting occultations. Apparently, this asymmetry reflects a natural higher attention of the observers to the dark part of the limb of an evening Moon. To exclude this observational selection effect, we consider in Fig.~\ref{Fig14}b the occurrence probability $P = n/N$, which demonstrates significant increase in the area of the poles. This increase is related with a geometry effect of the grazing occultations in view of the fact that all stellar occultation trajectories are directed almost perpendicularly to the lunar rotation axis, i.e., pass form the east to west side of the limb. Correspondingly, the total number $N$ of occultations, which fall into the sliding window $\delta \phi = 20^{\textrm{o}}$, centered at a given $\phi$, is $N \varpropto |\sin \phi |$. Therefore, $P$ increases at $\phi \rightarrow 0$ and/or $\phi \rightarrow 180^{\textrm{o}}$.

\begin{figure}
   \centering
   \includegraphics[width=0.40\textwidth,clip=]{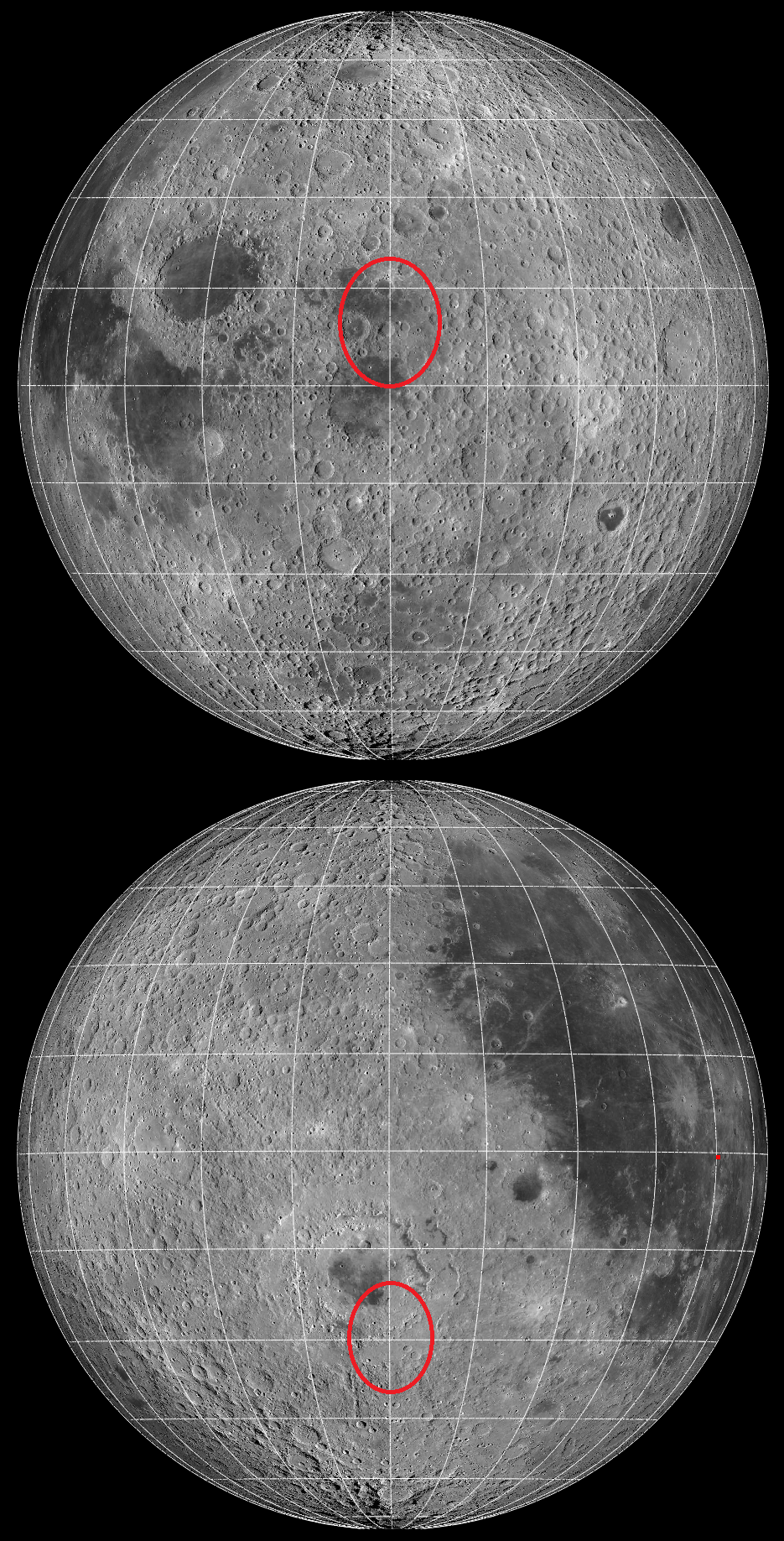}
   \caption{Regions (red ellipses) of an increased occurrence probability of the listed in the LOA long-lasting occultation events seen in  Fig.~\ref{Fig14}c. The top and bottom panels show the eastern and western hemispheres of the Moon, respectively. In both cases, North pole is at the top.}
              \label{Fig15}
    \end{figure}

To reduce the value of $P$ to its equatorial figure at $\phi = 90^{\textrm{o}}$ and $270^{\textrm{o}}$, we apply a correction factor and consider the corrected probability $P_{\textrm{c}} \equiv P |\sin \phi|$, which is shown in Fig.~\ref{Fig14}c. As result, two approximately equal peaks appear at $\phi \approx 120^{\textrm{o}}$ and $\phi \approx 280^{\textrm{o}}$. The corresponding latitudes of this increased dust activity on the lunar surface are $30^{\textrm{o}} \pm 10^{\textrm{o}}$S and $10^{\textrm{o}} \pm 10^{\textrm{o}}$N, respectively. The relevant longitudes are limited by the libration in longitude, i.e., $90^{\textrm{o}}\pm 8^{\textrm{o}}$W and $90^{\textrm{o}}\pm 8^{\textrm{o}}$E respectively. This finally points at the regions, marked in Fig.~\ref{Fig15}. It is remarkably that both regions include Mare Orientale, Mare Smythii and Mare Marginis. Such sites of an ancient sea volcanism (Whitten \& Head 2013) are known with respect of the lunar outgassing, which could lift up the surface dust (Meslin et al. 2020).

\begin{figure}
   \centering
   \includegraphics[width=0.40\textwidth,clip=]{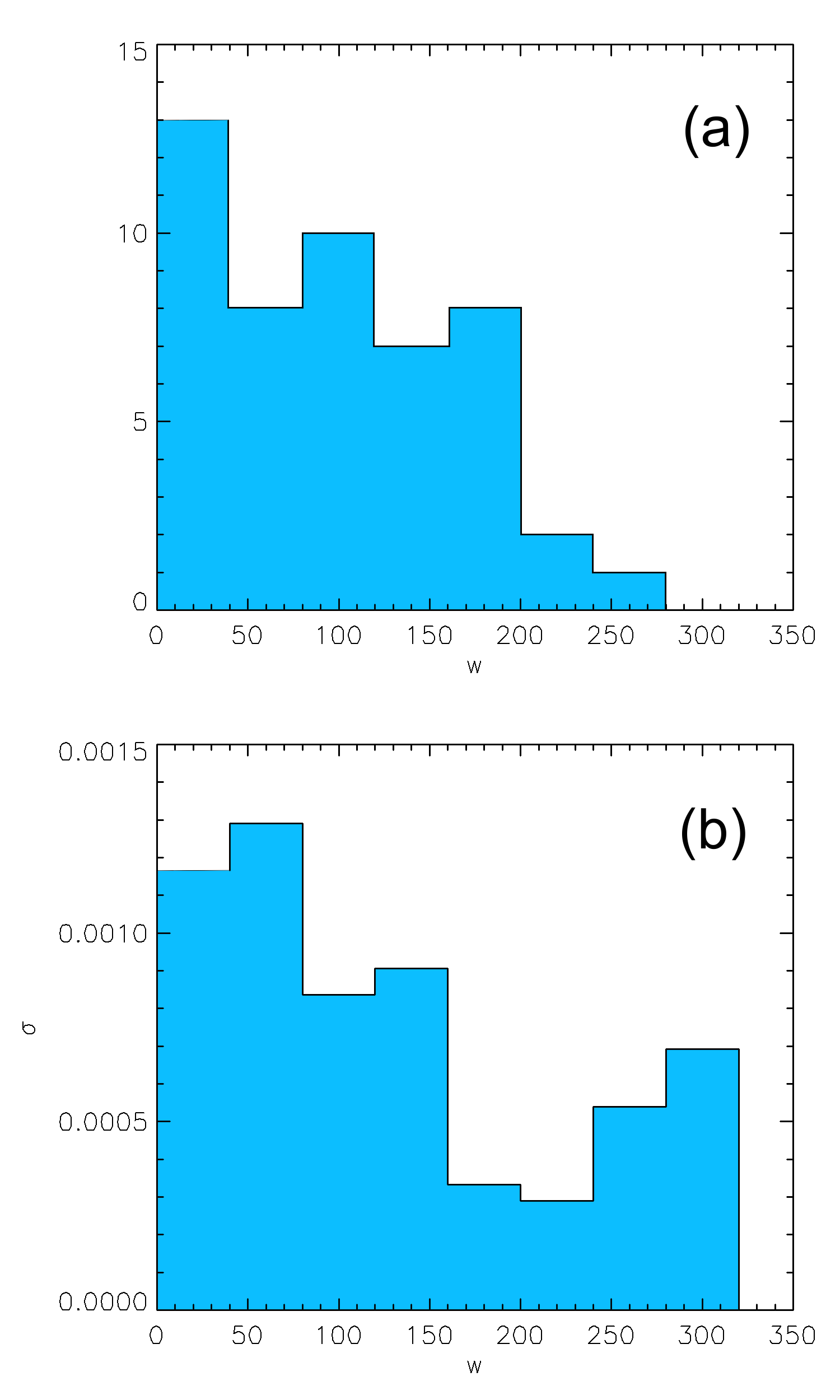}
   \caption{Distributions of the observed dust clouds' manifestations versus the daily total sunspot number $W$: \textbf{(a)} the histogram of all available manifestations in Tables 1-4 (the duplicates of the same dates are excluded); \textbf{(b)} the distribution of occurrence probability $P$ for the observed long-lasting occultation events from the LOA with $\tau \geq 0.1$ s .}
              \label{Fig16}
    \end{figure}

Since the solar wind plasma flow and UV radiation are the common attributes of the modern view of levitating lunar dust (e.g., Mishra and Bhardwaj  2019, Yeo et al. 2021), the connection between the real dust clouds (probed ba the stellar occultations) and solar activity deserves a special attention. The daily total sunspot number $W$ was included in our Tables 1-4 and is summarized in Fig.~\ref{Fig16}a. For comparison, we present in Fig.~\ref{Fig16}b the occurrence probability $P$ versus $W$ for the long-lasting ($\tau \geq 0.1$ s) occultations from the LOA. Both these plots demonstrate the tendency of the lunar dust clouds to appear during the periods of the low level solar activity. This fact contradicts an intuitive theoretical expectation of an increased density of the levitating dust during the periods of high solar activity, when the electric charging of dust particles is intensified. At the same time, the decreasing probability of dust clouds with the increasing solar activity might be result of a pick-up and removal of the lunar dust by the solar wind plasma flow. Such a blowing away process is confirmed in the data from the LADEE space probe as the presence of dust in the lunar "sodium tail", i.e., in a wake in the solar wind behind the Moon (Wooden et al. 2014, Cook et al. 2015).

\section{Dust risk for a space mission}

The published hitherto estimates of the dust concentration around the Moon are surprisingly dispersed in a wide range. In particular, McCoy (1974, 1976) based on the Apollo data about dust particles with diameters of $d=0.1 \ \mu\textrm{m}$, estimated their concentration varying from $n_{d}=10 \ \textrm{m}^{-3}$ at the altitude of $h=100$ km up to $n_{d}=10^{5} \ \textrm{m}^{-3}$ at $h=1$ km. Nevertheless, the dust detector of the LADEE mission found at $h=10$ km the density of $0.1 \ \mu\textrm{m}$ particles only at the upper limit of its sensitivity, that is approximately two orders of magnitude below the Apollo estimates. At the same time, the concentration of dust particles with $d \gtrsim 0.3 \ \mu\textrm{m}$ at the low altitudes of up to 50 km appeared to be only $n_{d} \lesssim 4 \times 10^{-3} \textrm{m}^{-3}$ (Hor\'{a}nyi et al. 2015).

Moreover, the \textit{Lunar Reconnaissance Orbiter} and \textit{Clementine} missions did not observe any forward light scattering from the dust above the lunar horizon (Feldman et al. 2014; Glenar et al. 2014). Accordingly, the dust clouds are not considered among the risks for the lunar missions. However, the archived data on the Earth-based observations of the phenomena related with the lunar dust, discussed in this paper, point at certain spatial and temporal variability of the dust formations around the Moon, calling for reevaluation of the associated potential technological risks.

\subsection{Risk of large-scale dust formations}

The long-living global dust cloud could be a most damaging phenomenon, which could slow down spacecraft in a low circumlunar orbit, resulting in its crash on the Moon. The relatively rarely seen annular Moon is an observational signature of this dangerous large-scale dust formation. To estimate the time $T_{c}$, during which a spacecraft could be decelerated inside the dust cloud with a consequent crash on the Moon, one needs an estimation of the dust concentration $n_{d}$.

According to Boltzmann's statistic theory, the probability $p$ of a dust grain to have a given potential energy $\varepsilon \varpropto z$ is $p \varpropto \exp(-\varepsilon/\langle \varepsilon \rangle)$, where $z$ is an altitude above the lunar surface, and $\langle \varepsilon \rangle$ is an average energy per a grain. Correspondingly, the dust number density versus height looks as follows:
\begin{equation}
 n_{d}(z) = n_{\textrm{o}} \exp(-z/h),
 \label{Eq3}
\end{equation}
where $n_{\textrm{o}}$ is concentration of dust at the base level $z=0$, and $h$ is the scale height. In practice, this approach is a standard approximation for the altitude distribution of lunar dust (see, e.g., Zook \& McCoy 1991).

There is a useful relation, followed directly from Eq. (\ref{Eq3}), between the dust column concentration $N_{d}$ above the lowest visible level $H$ of the dust cloud and the scale height:
\begin{equation}
 N_{d} = \int^{\infty}_{H} n_{d}(z) dz = n_{\textrm{o}} h \exp(-H/h) \equiv n_{\textrm{H}} h.
 \label{Eq4}
\end{equation}
We introduced in Eq. (\ref{Eq4}) the concentration of dust $n_{\textrm{H}}$ at the edge of the Moon shadow, i.e., at the point $A$  in Fig.~\ref{Fig09}, which is the lowest visible illuminated point on the line of sight $AB$, tangential to the lunar limb. The corresponding transition to the surface concentration is $n_{\textrm{o}} = n_{\textrm{H}} \exp(H/h)$.

The parameters $N_{d}$ and $h$ can be estimated from observations. In particular, the scattered sunlight flux from the unit cross-section of the column $AB$ in Fig.~\ref{Fig09} along the line of sight, i.e., the brightness $B_{aM}$) of an annular Moon, is
\begin{equation}
 B_{aM} = E_{\textrm{S}} \sigma_{s} \Phi(\theta) N_{d},
 \label{Eq5}
\end{equation}
where $E_{\textrm{S}}$ is the irradiation from the sunlight, $\sigma_{s}$ is the scattering cross section of a dust grain, and $\Phi(\theta)$ is the phase angle function of the scattering.

To be noticeable on Earth, this brightness must be $B_{aM} = K B_{\textrm{ash}}$, where $1.1 \lesssim K \lesssim 10$ is a reasonable excess factor to ensure visual observation, and $B_{\textrm{ash}}$ is a brightness of the lunar ashen glow, or the luminosity of an earthlight reflected from the dark side of the Moon, which can be expressed as follows:
\begin{equation}
 B_{\textrm{ash}} \approx E_{\textrm{S}} A_{\textrm{E}} A_{\textrm{M}} \left(\frac{R_{\textrm{E}}}{a_{\textrm{M}}}\right)^2.
 \label{Eq6}
\end{equation}
Here $A_{\textrm{E}} = 0.434$ is the geometric albedo of the Earth in V-band (Mallama et al. 2017), $A_{\textrm{M}} = 0.1362$ is the lunar geometric albedo (Grant 2008), $R_{\textrm{E}} = 6371$ km is the mean radius of the Earth, and $a_{\textrm{M}} = 384399$ km is the lunar orbit semi-major axis.
Therefore, the above formulated condition for the brightness of an annular Moon (in order this phenomenon is visible on the Earth) after substitution of Eqs. (\ref{Eq5}) and (\ref{Eq6}) in it, gives the corresponding dust column concentration:
\begin{equation}
 N_{d} \approx K \frac{A_{\textrm{E}} A_{\textrm{M}}}{\sigma_{s} \Phi(\theta)} \left(\frac{R_{\textrm{E}}}{a_{\textrm{M}}}\right)^2,
 \label{Eq7}
\end{equation}
Finally, taking Eq. (\ref{Eq4}) into account, we obtain from Eq. (\ref{Eq7}) the dust concentration
\begin{equation}
 n_{\textrm{H}} \approx K \frac{A_{\textrm{E}} A_{\textrm{M}}}{h \sigma_{s} \Phi(\theta)} \left(\frac{R_{\textrm{E}}}{a_{\textrm{M}}}\right)^2,
 \label{Eq8}
\end{equation}
at the edge of the Moon's shadow, i.e., at point $A$ in Fig.~\ref{Fig09}.

The values of $\sigma_{s}$ and $\Phi(\theta)$ in Eq. (\ref{Eq8}) are calculated using the Mie scattering  theory\footnote{\url{https://omlc.org/calc/mie_calc.html}}, for the wavelength $\lambda = 0.5 \ \mu\textrm{m}$ at the maximum of sunlight spectrum, assuming the dust grains as transparent glass spheres with the refractive index 1.5, having a zero imaginary part. The phase function $\Phi(\theta)$ is normalized so that its integral over $4 \pi$ steradians is unit. Since the average angular distance between the Moon and the Sun, $\beta$, according to Table 4, is $\langle \beta \rangle = 36^{\textrm{o}}$, we accepted in $\Phi(\theta)$ the phase angle $\theta \approx 180^{\textrm{o}} - \langle \beta \rangle = 144^{\textrm{o}}$.

To find an expression for the dust scale height $h$, we consider the grain potential energy $\varepsilon = g_{\textrm{eff}} \ m_{d} \ z$, used for the derivation of Eq. (\ref{Eq3}), which is a function of the grain mass $m_{d}$, and effective surface gravity $g_{\textrm{eff}}$. The latter includes also a contribution of the lunar electric field. Correspondingly, the scale height is related to the grain mass, as $h \varpropto m_{d}^{-1}$. Hence, one can accept the following expression:
\begin{equation}
 h = \frac{C}{m_{d}},
 \label{Eq8b}
\end{equation}
where $C$ is a constant.

Let us estimate the constant $C$ in Eq. (\ref{Eq8b}) by considering, at first, the dust grains of a minimal mass, assuming them as spheres with a minimal diameter $d_{\textrm{min}}=0.1 \ \mu\textrm{m}$ and density $\rho_{d} = 2500 \ \textrm{kg \ m}^{-3}$ which corresponds to a spherical impact glass (see for its justification in McCoy 1976; Glenar et al. 2020). This altogether gives us the dust grain minimal mass of $m_{\textrm{min}}= (4/3) \pi \rho_{d} (d_{\textrm{min}}/2)^3 = 1.3\times10^{-18}$ kg. Such a finest dust should naturally have the maximal scale height $h_{max}$, which can be estimated from observations of the annular Moon phenomena. In particular, averaging the listed in Table 4 minimal heights $H$ of the illuminated dust, estimated with Eq. (\ref{Eq1}), which enable the light ring closure around dark side of the lunar limb, providing $\gamma = 90^{\textrm{o}}$, gives $\langle H \rangle = 108$ km. On the other hand, the fact of visibility of an annular Moon in the sky with a naked eye gives approximately the same height of 116 km, which corresponds to an eye resolution of about 1 arc minute. Moreover, such an average height of a large-scale dust cloud was confirmed by Apollo astronauts, who registered visually and photographically the diffuse glow of the illuminated dust at the orbital altitudes of 100-120 km above the lunar surface (McCoy 1976). The space probe LADEE also registered a continuous dust cloud around the Moon at similar altitudes of 100-150 km (Hor\'{a}nyi et al. 2020). Therefore, based on the said above, we finally adopt as an average scale height of a global circumlunar dust cloud, the value $h = h_{\textrm{max}} = 116$ km and obtain from Eq. (\ref{Eq8b}) the constant $C = m_{\textrm{min}} h_{\textrm{max}} = 1.52\times10^{-13}$ kg m.

The shorter scale heights of dust formations range, according to Apollo reports, from 5 to 20 km (Zook \& McCoy 1991).
These estimates are consistent with the outcomes of considered in sections 2.1 and 2.2 Earth-based observations, which altogether give an average dust layer thickness $\langle H \rangle$ between 6 km (Table 3) and 23 km (Table 1). Within the above adopted approximation of the dust grain shape and density, the Eq. (\ref{Eq8b}) with the same value of constant $C$, as that estimated for the high altitude annular Moon dust clouds, yields for the lower dust clouds, a larger diameter $d \sim 0.2$ $\mu\textrm{m}$ of the levitating dust grains, which corresponds the smaller scale height $h = h_{\textrm{min}} = 14.5$ km.

\begin{figure}
   \centering
   \includegraphics[width=0.50\textwidth,clip=]{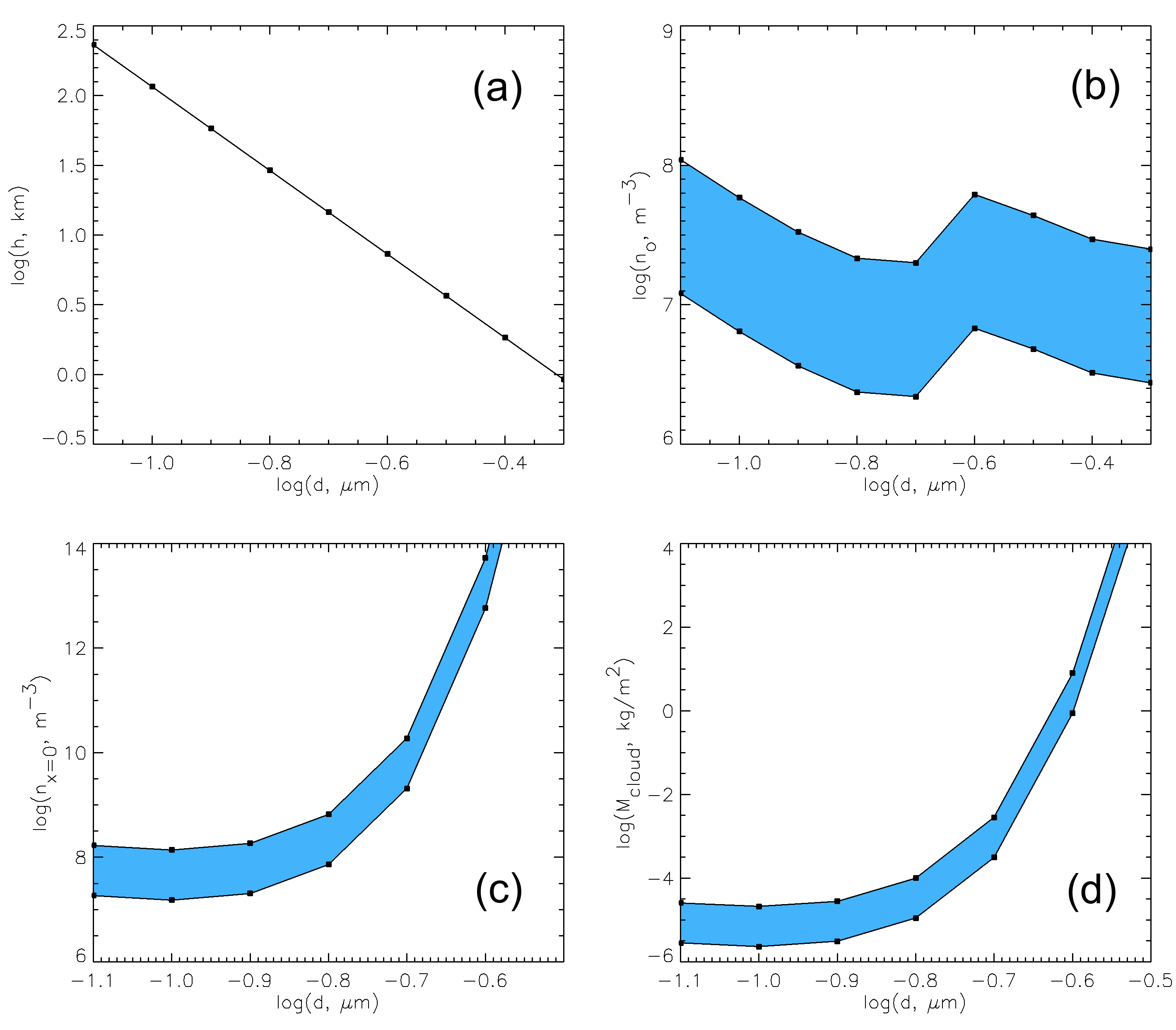}
   \caption{Estimates of the typical dust cloud parameters, providing an annular Moon phenomenon, as functions of the dust grain accepted diameter $d$: \textbf{(a)} the scale height $h$ according to Eq. (\ref{Eq8b}); \textbf{(b)} the concentration $n_{\textrm{\textrm{H}}}$ at the lowest visible illuminated dust level $H \equiv \langle H \rangle = 108$ km according to Eq. (\ref{Eq8}); \textbf{(c)} the surface dust concentration $n_{\textrm{o}}=n_{\textrm{H}} \exp(H/h)$; \textbf{(d)} the dust column mass density $M_{\textrm{cloud}} = n_{\textrm{o}} \  h  \ m_{d} $ per unit area of lunar surface. The colored ranges correspond to the confidence intervals between the curves calculated with the adopted minimal and maximal values of the visibility factor $1.1 \lesssim K \lesssim 10$.}
              \label{Fig17}
\end{figure}

Figure~\ref{Fig17} shows the parameters of dust cloud revealed with the above described model. As it can be seen in Fig. ~\ref{Fig17}b, the annular Moon phenomenon requires at a typical height $H \equiv \langle H \rangle = 108$ km of the lowest visible level of the dust cloud (according Table 4) the concentration of dust $n_{\textrm{H}} \sim 10^{7} \textrm{m}^{-3}$. Correspondingly, the surface values of dust concentration for the grain size $d \lesssim 0.16 \ \mu\textrm{m}$ ($\log{0.16} \approx -0.79$) appear $n_{\textrm{o}} \sim 10^{8} \ \textrm{m}^{-3}$, as shown in Fig.~\ref{Fig17}c. This is $\sim 10^{3}$ times higher than an estimate based on the Apollo measurements (McCoy 1976). However, such a great difference between the estimates of surface dust concentration does not seem impossible in view of even greater discrepancies of $>10^{3}$ times, obtained for the column dust concentration, using the data from \textit{Apollo} (McCoy 1976) and \textit{Clementine} (Glenar et al. 2014). The dust column mass density $M_{\textrm{cloud}} = \int^{\infty}_{0} n_{d}(z) m_d dz = n_{\textrm{o}} \  h  \ m_{d}$ is minimal for the dust grain size of $d \approx 0.1 \ \mu\textrm{m}$ arguing for the finest dust with its total mass $\sim 10^{-5}$ kg per 1 $\textrm{m}^{2}$.

Now, let us estimate the time $T_c$, during which a spacecraft with a mass $M_{sc}$ and cross-section $S$, flying at a low circular orbit with velocity $V$ inside an extended (large-scale) circumlunar dust cloud, would loss its orbital momentum due to collisions with the levitating dust particles and crash on the Moon. For simplicity, we assume no mass change for the spacecraft ($M_{sc}=const$), and take the velocity of dust particles to be much less, as compared with the spacecraft orbital speed $V$. A spacecraft energy loss rate due to collisions with the dust is
\begin{equation}
 \varepsilon = - (\frac{1}{2} m_{d} V^{2}) (n_{d} S V) = - \frac{1}{2} m_{d} n_{d} S V^{3},
 \label{Eq9a}
\end{equation}
were $m_{d}$ and $n_{d}$ are the mass and local density of dust particles, respectively. Since the total energy of spacecraft is $E = \frac{M_{sc} V^2}{2} - \frac{G M_M M_{sc}}{r}$, where $r=R+z$ is a selenocentric distance of spacecraft (measured from the Moon center) with $R=1737$ km, $M_M = 7.35 \times 10^{22}$ kg is mass of the Moon, and $G = 6.67 \times 10^{-11} \textrm{m}^{3} \textrm{kg}^{-1} \textrm{s}^{-2}$ is the gravitational constant, we obtain from the energy conservation law that
\begin{equation}
 \varepsilon = \frac{dE}{dt} = M_{sc} V \frac{dV}{dr} \frac{dr}{dt} + \frac{G M_M M_{sc}}{r^{2}} \frac{dr}{dt}.
 \label{Eq9b}
\end{equation}

For the quasi-circular orbits, a spacecraft orbital velocity is $V(r) \approx \sqrt{G M_{M}/r}$, so that $dV/dr = - V/(2 r)$. After substitution of these expressions in Eq. (\ref{Eq9b}), we obtain that
\begin{equation}
 \frac{dr}{dt} = \frac{2 r^{2} \varepsilon}{G M_M M_{sc}} = \frac{2 \varepsilon}{g(r) M_{sc}},
 \label{Eq9c}
\end{equation}
where $g(r) = G M_{M}/r^{2}$ is the lunar gravity. Finally, taking Eq. (\ref{Eq9a}) into account, the crash time $T_{c}$ of a spacecraft falling from its initial circular orbit at an altitude $z=Z_0$ can be found as
\begin{equation}
 T_{c} = \frac{M_{sc}}{2} \int_{R+Z_0}^{R} \frac{g(r) dr}{\varepsilon} = - \frac{M_{sc}}{m_{d} S} \int_{R+Z_0}^{R} \frac{g(r) dr}{V^{3}(r) n_{d}(r)}.
 \label{Eq10}
\end{equation}
The distribution of dust concentration $n_{d}(r)$ here is defined by Eq. (\ref{Eq3}) after substitution of $z=r-R$ with the base concentration $n_{\textrm{o}}$ at the lunar surface ($r=R$), i.e., at $z=0$ defined by Eq. (\ref{Eq8}) with the relation $n_{\textrm{o}} = n_{\textrm{H}} \exp(H/h)$ taken into account.

Since the dust concentration $n_{d}$ is anyway defined only approximately, due to using in Eq. (\ref{Eq8}) of the visibility factor $1.1 \lesssim K \lesssim 10$, which ensures sufficient brightness of the annular Moon phenomenon, influencing in its turn the value of $n_{\textrm{o}}$, it makes no sense to solve precisely the integral in Eq. (\ref{Eq10}), which we replace with an approximated estimate. In particular, for a low quasi-circular orbit of a spacecraft at an altitude $z = (r - R) \ll R$, we substitute in Eq. (\ref{Eq10}) instead of $g(r)$ and $V(r)$ the approximately constant values of the lunar surface gravity $g_M =\frac{G M_M}{R^2} \approx 1.622 \ \textrm{m}/\textrm{s}^{2}$ and an orbital speed $V_M = \sqrt{G M_{M}/R} \approx 1.68 \ \textrm{km/s}$. It is worth mentioning that for a typical value of low-altitude spacecraft parking orbit $Z_0 = 100$ km, the variation of orbital speed $V(r) = \sqrt{G M_{M}/(r)}$ in the altitude range of $R \leq r \leq R+ Z_0$ is $|(\partial V(r)/\partial r) Z_0| = V(r)/(2 r) Z_0 \sim 0.5 V_M Z_0 / R \sim 0.03 V_M$. Therefore, neglecting by variability of $V(r)$ under the integral in Eq. (\ref{Eq10}) and assuming $V(r) = V_M$ is well justified in view of a one-order of magnitude uncertainty of the $n_{d}$ value.

As a result, Eq. (\ref{Eq10}) yields the following estimate of the crash time:
\begin{equation}
 T_{c} \approx \frac{g_M M_{sc} h}{m_{d} n_{\textrm{o}} S V_{M}^{3}} \left[ \exp \left( \frac{Z_0}{h} \right) - 1 \right].
 \label{Eq11}
\end{equation}

To calculate finally the value of $T_c$, we take in Eq. (\ref{Eq11}), along with of the already defined above numeric parameters, the scale height $h$ according to Eq. (\ref{Eq8b}) and the spacecraft mass $M_{sc} = \rho_{sc} S^{3/2}$ with the density $\rho_{sc}=100$ kg m$^{-3}$ (like for the \textit{Vikram} or \textit{Beresheet} missions). The effective cross-section of spacecraft was taken as $S=1$ \textrm{m}$^{2}$. The latter simplifies the conversion of the obtained $T_c$ to the cases of spacecraft with other cross-sections just by applying of the factor $\sqrt{S/1 \ \textrm{m}^{2}}$.

\begin{figure}
   \centering
   \includegraphics[width=0.50\textwidth,clip=]{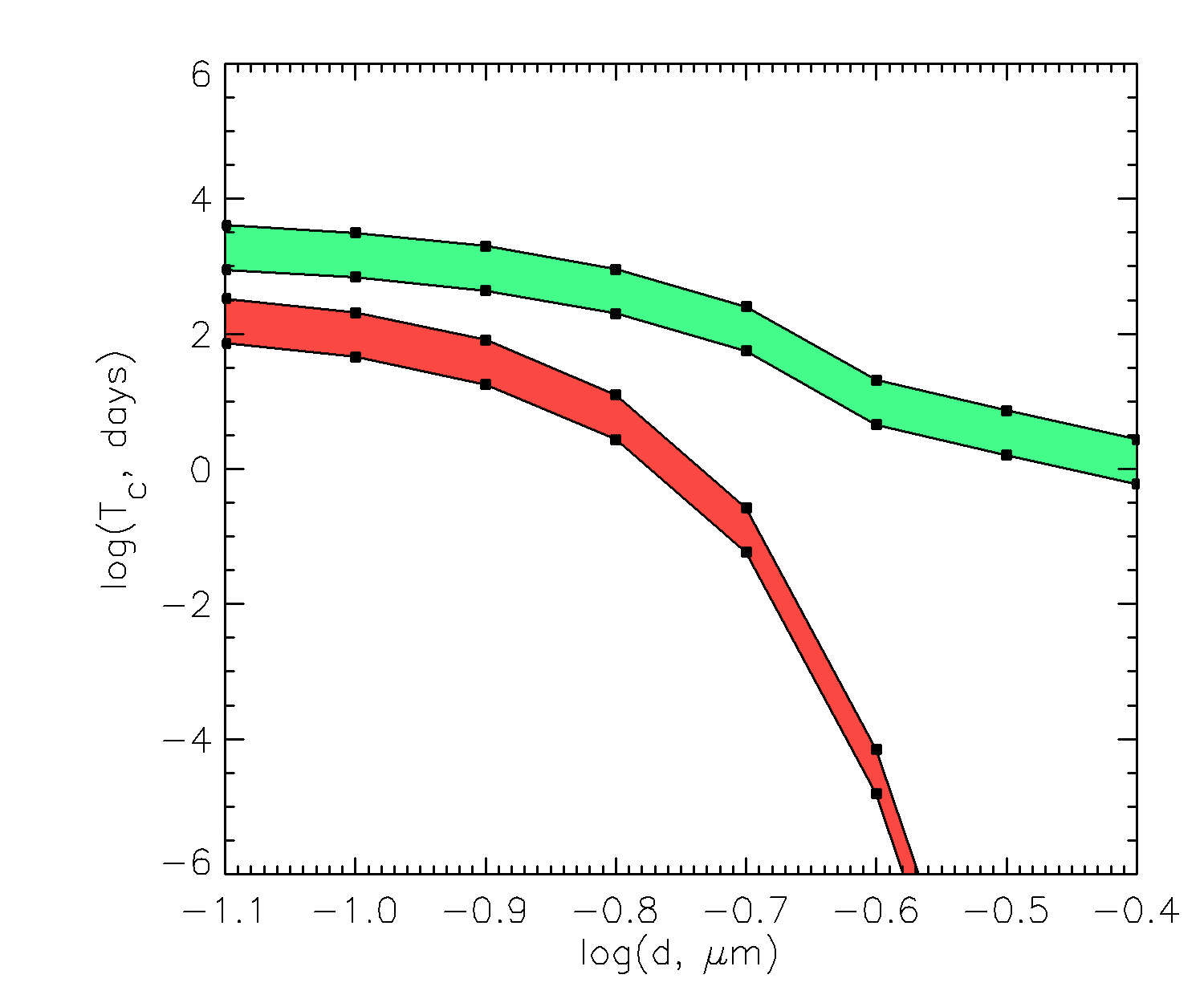}
   \caption{A spacecraft crash time $T_{c}$, according to Eq. (\ref{Eq11}), for the initial altitudes $Z_0=100$ km (green) and for $Z_0=10$ km (red). The colored ranges correspond to the confidence intervals between the curves calculated with the adopted minimal and maximal values of the visibility factor $1.1 \lesssim K \lesssim 10$.
   }
              \label{Fig18}
    \end{figure}

Figure~\ref{Fig18} shows the results of our calculations of the crash time $T_{c}$ of a spacecraft, using Eq. (\ref{Eq11}). The cases with two initial altitudes of a spacecraft, $Z_0 = 100$ km and $Z_0 = 10$ km, were considered. These correspond to a typical lunar orbiter parking and a low-altitude pre-landing orbits, respectively. One can see, that the lifetime of a spacecraft at the parking orbit ($T_{c} \gg 1$ day) in the case of a sufficiently fine dust cloud with $d < 0.2 \ \mu\textrm{m}$ ($\log{0.2} \approx - 0.70$) is much longer than the timescale of an annular Moon event ($\lesssim 1$ day). Therefore, the parking orbits with $Z_0 \gtrsim 100$ km are relatively safe in the case of fine dust, but the safety problem might become an issue for the larger dust grain sizes, e.g., for $d \gtrsim 0.4 \ \mu\textrm{m}$ ($\log{0.4} \approx - 0.40$). At the same time, a dramatic dust effect takes place at the low-altitude pre-landing orbits with $Z_0 \lessapprox 10$ km in the case of a moderate size of dust grains $d \gtrsim 0.2 \ \mu\textrm{m}$. In this case, the crash time of a spacecraft, according to Figure~\ref{Fig09}b, is $T_{c} \lesssim 0.12$ day or less than one orbit. Although Eq. (\ref{Eq11}) gives a rather rough approximation for such a short crash time, its result can be still used as an alarming indicator.

\subsection{Risk of short-scale dust formations}

Now let us consider the spacecraft potential danger from the short-scale dust formations, such as the impact plumes, manifested in the long stellar occultations (see Sect. 3.2). The extinction $\Delta F$ of the light flux $F$ in a thin dusty layer of a plume with a thickness $\Delta x$ along the direction of view is
\begin{equation}
 \Delta F = -F \sigma_{s} n_{d} \Delta x.
 \label{Eq13}
\end{equation}

\begin{figure}
   \centering
   \includegraphics[width=0.4\textwidth,clip=]{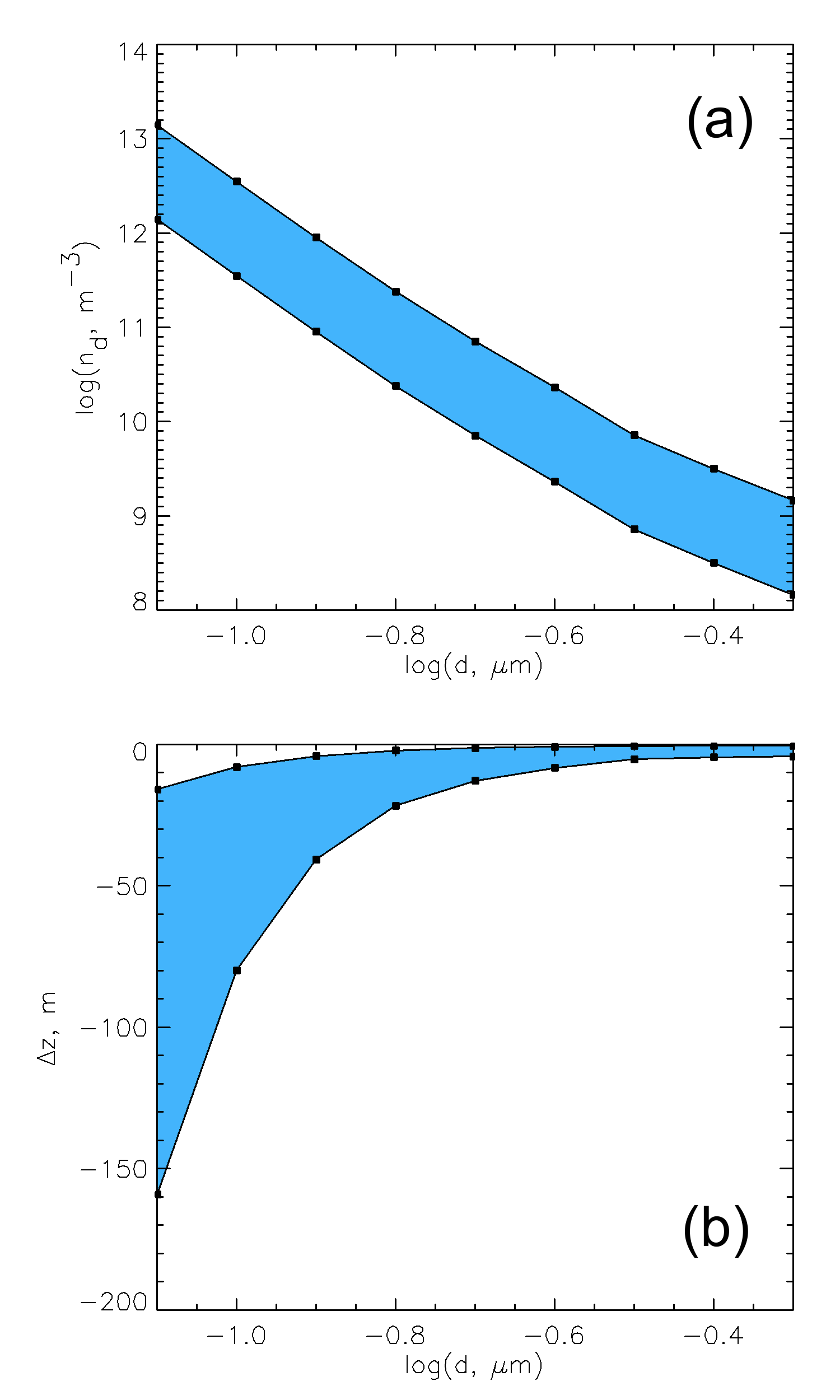}
   \caption{Effect of a typical impact plume with $L = 1$ km, manifested in the long stellar occultations: \textbf{(a)} an average value of dust concentration $n_{d}$ in the plume for an assumed reasonable range of its optical thickness $0.1 \lesssim \tilde{\tau} \lesssim 1$ providing detectability of the phenomenon; \textbf{(b)} the decrease of a spacecraft altitude $\Delta z$ during a single flyby through the plume, according to Eq. (\ref{Eq16}). The colored ranges correspond to the adopted interval of $\tilde{\tau}$. The same parameters of spacecraft and numeric values were applied, as for the graphs in Fig.~\ref{Fig18}.}
              \label{Fig19}
    \end{figure}

After integration over the whole plume scale $L$, we obtain the known expression
\begin{equation}
 F = F_{o} \exp(-\tilde{\tau}),
 \label{Eq14}
\end{equation}
where $F_{\textrm{o}}$ is the incoming stellar light flux, and $\tilde{\tau} \equiv \sigma_{s} n_{d} L$ is the optical thickness.
Correspondingly, one can estimate an average dust concentration as
\begin{equation}
 n_{d} = \frac{\tilde{\tau}}{\sigma_{s} L},
 \label{Eq15}
\end{equation}
Assuming $L = 1$ km, according to Fig.~\ref{Fig10}b and taking the value of $\tilde{\tau}$ in a reasonable range of the extinction detectability, e.g., $0.1 \lesssim \tilde{\tau} \lesssim 1$, using Eq. (\ref{Eq15}), we estimate the value of $n_{d}$ (see in Fig.~\ref{Fig19}a).

A spacecraft single flyby through such a dust plume during the corresponding time $\Delta t = L/V$ will result, according to Eq. (\ref{Eq9c}), in the decrease of spacecraft's altitude, $\Delta z \equiv (dr / dt) \Delta t$, due to the energy loss rate $\varepsilon$ defined by Eq. (\ref{Eq9a}):
\begin{equation}
 \Delta z = \frac{dr}{dt} \Delta t = - \frac{m_{d} n_{d} S L V^{2}}{g_M M_{sc}}.
 \label{Eq16}
\end{equation}

Figure~\ref{Fig19}b shows the calculated with Eq. (\ref{Eq16}) values of $\Delta z$. One can see that the plume effect is negligible ($|\Delta z| < 10$ m) for $d \gtrsim 0.16 \ \mu\textrm{m}$ ($\log{0.16} \approx - 0.79$) for a suborbital vehicle (lander) at the low altitudes ($\sim L=1$km). However, it becomes more and more dangerous ($|\Delta z| > 100$ m) for the smaller dust grains ($d \lesssim 0.1 \ \mu\textrm{m}$) and correspondingly higher cloud concentrations ($n_{d} \gtrsim 5 \times 10^{11}$m$^{-3}$).

\section{Conclusions}

The available historical material, used along with the more recent observational data in the present work is practically unknown to the modern planetary community and, therefore, still remains unexplored. Nevertheless, it contains important data on potentially dangerous circumlunar dust phenomena, which seem to be overlooked with regard of recent space missions. Our review sheds a light on such phenomena revealing several important facts, regarding the behavior and origin of dust formations around the Moon and their potential danger for the space missions.

In particular, we show that the manifestations of the lunar dusty envelope have been repeatedly observed over the past three centuries, but they were incorrectly interpreted in terms of the lunar dense atmosphere. Practically all types of the possible optical effects from the lunar dust clouds were observed (Sect. 2), including the lunar horizon glow (see Table 3 and 4), which discovery is erroneously assigned to the landers and astronauts.

More significantly, the ground-based observers have seen a much grandiose dust phenomenon than those recorded by the lunar missions. Not only individual dust ejections, but the large-scale global dust clouds have been observed and described many times as the phenomenon of an annular Moon (Table 4). The facts of the ground-based observations of such phenomena indicate an incommensurable dust concentration, as compared with the data provided by modern, but limited in time and location, space missions. Apparently, the rarity of an annular Moon phenomenon is a reason for its non-detection in-situ.

Nevertheless, the investigated statistics of dust manifestations, observed from the Earth, correlates with the mainstream viewpoint on the origin of  the lunar dust exosphere, as a result of the meteoroid impacts (Fig.~\ref{Fig10} and \ref{Fig12}).

Over many years, the lunar outgassing was discussed as another source of the levitating dust. In this regard, our discovery of the solar tide related half-month (14.77 days) periodicity in the dust cloud appearance (Fig.~\ref{Fig13}), as well as their concentration in the regions of an ancient vulcanism (Fig.~\ref{Fig15}) appear of great interest.

The revealed tendency of dust clouds to be observed during the periods of low solar activity (Fig.~\ref{Fig16}) raises a question on possible entrainment of the dust by solar wind. In general, the interaction between lunar dust and solar wind appear of interest in the context of the observed anomalies in the transit photometry of exoplanets (Arkhypov et al. 2019, 2020, 2021), providing therefore an in-situ measurable environment for the investigation of such dusty plasma environments. The historical records on the lunar dust phenomena contribute to this studies as well.

Our estimations in Sect. 4 demonstrate that the global circumlunar clouds of a sub-micrometer dust are dangerous for the space vehicles on the pre-landing low circular orbits at the altitudes below 10 km. The relevance of this result might be related with the crashes of the lunar landers \textit{Vikram} and \textit{Beresheet}, after the reported sudden malfunctions at the altitudes $2.1$ and $<15$ km, respectively.

The increasing humankind space activity around the Moon, including the plans of manned missions, needs a particular attention to the risks in light of the addressed in the present work reports on various, yet poorly explored, circumlunar dust phenomena.

\section{Acknowledgements}

This research used the \textit{Lunar Occultation Archive} (Herald et al. 2022). The authors acknowledge the support by the projects I2939-N27 and S11606-N16 of the Austrian Science Fund (FWF) during the period of the development of the analysis techniques, applied in the present study.

\section{References}

  Arkhypov, O.V., Khodachenko, M.L., Hanslmeier, A. 2019, A\&A, 631, A152

  Arkhypov, O.V., Khodachenko, M.L., Hanslmeier, A. 2020, A\&A,  638, A143

  Arkhypov, O.V., Khodachenko, M.L., Hanslmeier, A., 2021, A\&A, 646, A136

  Arsyukhin E. 1994, Zemlya i Vselennaya, 2, 78

  Barabashev N.P. 1915, Izvestiya ROLM, 4, 5(17), 216

  Bickerdike. 1867, English Mechanic, 4(95), 277

  Brock H. 1969, Sky and Telescope, 37, 122

  Burr T.W. 1863, \mnras, 23, 221

  Cameron W.S. 1977, Physics of the Earth and Planetary Interiors, 14, 194

  Cameron W.S. 1978, Lunar transient phenomena catalog. NSSDC/WDC-A-R\&S 78-03, Greenbelt: NASA

  Challis J. 1857, \mnras, 17, 136

  Cheng A.F., Stickle A.M., Fahnestock E.G., Dotto E., Della Corte V., Chabot N.L. \& Rivkin A.S. 2020, Icarus, 352, 113989

  Cook A. M., Wooden D. H., Colaprete A., Glenar D. A., Stubbs T. J. 2015, 46th Lunar and Planetary Science Conference, LPI Contribution 1832, 2147

  Corliss W.R. 1979, Mysterious Univers: a handbook of astronomical anomalies. Glen Arm: The Sourcebook Project, 157.

  Corliss W.R. 1985, The Moon and the planets: a catalog of astronomical anomalies. Glen Arm: The Sourcebook Project, 157.

  Crotts A.P.S., Hummels C. 2009, \apj, 707, 1506

  Dawes W.R. 1863, \mnras, 23, 221

  Deka R., Bora M.P. 2018, Physics of Plasmas, 25(10), 103704

  Denett F. 1877, English Mechanic, 25, 89

  Dollfus A. 2000, Icarus, 146(2), 430

  Dunn S. 1762, Phil. Trans., 52, 578

  Feldman P.D., Glenar D.A., Stubbs T.J., Retherford K.D., Randall G.G., Miles P.F., Greathouse T.K., Kaufmann D.E., Parker J.Wm., Alan S.S. 2014, Icarus, 233, 106

  Fitzpatrick J.A. 1944, Sky and Telescope, 3(9), 20

  Flammarion C. 1880, Astronomie populaire. Paris: G. Marpon, E. Flammarion, 241

  Flammarion C. 1882, L'Astronomie, 1, 188

  Flammarion C. 1890, l$'$Astronomie, 9(7), 256

  Florenski P.V., Chernov V.M. 1973, Astronom. Herald, 7(1), 38

  Florensky P.V., Chernov V.M. 1975, Astronom. Herald, 9(3), 194

  Florenskj P.V., Chernov V.M. 1994, Astronom. Herald, 28(4-5), 235

  Gheury M.E.J. 1913, The Observatory, 36, 268

  Glenar D.A., Stubbs T.J., Hahn J.M., Wang Y. 2014, J. Geophys. Res.: Planets, 119, 2548

  Glenar D.A., Stubbs T.J., Grava C., Retherford K.D. 2020, LPI Contribution, 2141, 5033

  Goddard J.F. 1856, \mnras, 17, 4

  Grant R. 1852, History of physical astronomy. London: H.G. Bohn, 3, 230

  Grant M. 2008, Applied Optics, 47(27), 4981

  Grove W.R. 1856, \mnras, 17, 3

  Haas W.H. 1942, J. Roy. Astronom. Soc. of Canada, 36, 361

  Haas W.H. 1944, JRASC, 38, 351

  Haas W.H. 1948, JALPO, 2(4), 2

  Haas W.H. 1950a, JALPO, 4(7), 7

  Haas W.H. 1950b, JALPO, 4(10), 8-9

  Haas W.H. 1955, JALPO, 9(11-12), 142

  Haas W.H. 1957, JALPO, 11(7-10), 118

  Herald D., Gault D. \& Iota team, 2022, Lunar Occultation Archive, VizieR On-line Data Catalog, VI/132C

  Hopkins B.J. 1883, Astronomical Register, 21, 140

  Hor\'{a}nyi M., Szalay J. R., Kempf S., Schmidt J., Gr\"{u}n E., Srama R. \& Sternovsky Z. 2015, Nature, 522(7556), 324

  Hor\'{a}nyi M., Bernardoni E., Carroll A., Hood N., Hsu S., Kempf S., et al. 2020, LPI Contribution, 2141, 5032

  Jacob W.S. 1848, \mnras, 8, 27

  Klein H.J. 1879, Sirius, 12(4), 132

  K\"{u}veler G., Klemm R. 1972, Sterne und Weltraum, 8-9, 238

  Laussedat, de Salicis, Mannheim, Bour, Girard, 1860, Compt. Rend. Acad. Sci. Paris, 51, 990

  Layard A.H. 1853, Discoveries in the ruins of Nineveh and Babylon. London: J. Murray, 607.

  Leeson C. 1863, \mnras, 24, 26

  Leavens P.A. 1944, Sky and Telescope, 3(4), 2

  Lenard Ph., Wolf M., Galle J. G., Geelmuyden H., Weinek L., F\'{e}nyi J. 1892, AN, 130(9-10), 145

  Lianghai X., Xiaoping Z., Lei L., Bin Z., Yiteng Z., Qi Y., Yongyong F., Dawei G. \& Shuoran Y. 2020,
                               Geophys. Res. Let., 47(23), e89593

  Loftus A.J. 1875, Nature, 12, 495

  Luizard M.P.E. 1933, L'Astronomie, 47, 276

  Mallama A., Krobusek B. \& Pavlov H. 2017, Icarus, 282, 19

  McCoy J.E., Criswell D.R. 1974, Geochimica et Cosmochimica Acta. Suppl. 5, 3, 2991

  McCoy J.E. 1976, Proc. Lunar Sci. Conf. 7th, 1087

  Meslin P. Y., He H., Kang Z. et al. 2020, 51st Lunar and Planetary Science Conference, LPI Contribution No. 2326, 1741

  Mishra S.K. and Bhardwaj A. 2019, ApJ, 884, 5

  Noble W. 1863, \mnras, 23, 251

  Noble W. 1875, \mnras, 36, 40

  Noble W. 1882, Knowledge, 2, 433

  Noble W. 1884, \mnras, 44, 261

  Pickering W.H. 1892, \aa, 11, 778

  Potter A.E., Zook H.A. 1992, Bulletin of the American Astronomical Society, 24, 1020

  Pratt H. 1884, l'Astronomie, 3(7), 270

  Reed G. et al. 1974, Nature, 247, 447

  Rennilson J.J. \& Criswell D.R. 1974, The Moon, 10, 121

  Rodgers J., Hammer J. 1878, Scientific American, 39, 385

  Simms W. 1857, \mnras, 17, 80

  Schroeter J.J. 1792, Phil. Trans. 82, 309

  Stolyarenko D. 1912, Izvestiya ROLM, 1(3), 26

  Stone E.J. 1889, \mnras, 50, 38

  Vertu J. 1863, \mnras, 23, 252

  Wates C.G. 1944, JRASC, 38, 171

  Whitten J.L., Head J.W. 2013, 44th Lunar and Planetary Science Conference, LPI Contribution No. 1719, 1247

  Wilkins P.H. 1951, JALPO, 5(9), 2

  Wooden D.H., Cook A., Colaprete A., Shirley M., Vargo K., Elphic R. C., Hermalyn B., Stubbs T. J., Glenar D. A. 2014. American Geophysical Union, Fall Meeting 2014, abstract id.P23C-4004

  Yeo L.H., Wang X., Deca J.,Hsu H.-W., Horanyi M. 2021, Icarus, 366, 114519

  Zook H.A., McCoy J.E. 1991, Geophys. Res. Let., 18(11), 2117

\end{document}